\documentclass[preprint]{JHEP3} % 10pt is ignored!
\pdfoutput=1

\usepackage{multicol,bbm,amsmath,url}
\usepackage{graphicx} %% REMOVE 'DRAFT' TO SEE FIGS

\newcommand\fverb{\setbox\pippobox=\hbox\bgroup\verb}
\newcommand\fverbdo{\egroup\medskip\noindent%
			\fbox{\unhbox\pippobox}\ }
\newcommand\fverbit{\egroup\item[\fbox{\unhbox\pippobox}]}
\newbox\pippobox
\renewcommand\d{\partial}
\renewcommand\O{\mathcal O}

\title{Boost Breaking in the EFT of Inflation}

\author{Luca V.\ Delacr\'etaz$^a$, Toshifumi Noumi$^b$ and Leonardo Senatore$^{a,c}$ \\
$\!^a$ Stanford Institute for Theoretical Physics, Stanford University, Stanford, CA 94305, USA\\
$\!^b$ Jockey Club Institute for Advanced Study,
Hong Kong University of Science and Technology, Hong Kong\\
$\!^c$ Kavli Institute for Particle Astrophysics and Cosmology,
SLAC and Stanford University, Menlo Park, CA 94025, USA\\
}
\abstract{

If time-translations are spontaneously broken, so are boosts. This symmetry breaking pattern can be non-linearly realized by either just the Goldstone boson of time translations, or by four Goldstone bosons associated with time translations and boosts. In this paper we extend the Effective Field Theory of Multifield Inflation to consider the case in which the additional Goldstone bosons associated with boosts are light and coupled to the Goldstone boson of time translations. The symmetry breaking pattern forces a coupling to curvature so that the mass of the additional Goldstone bosons is predicted to be equal to $\sqrt{2}H$ in the vast majority of the parameter space where they are light. This pattern therefore offers a natural way of generating self-interacting particles with Hubble mass during inflation. After constructing the general effective Lagrangian, we study how these particles mix and interact with the curvature fluctuations, generating potentially detectable non-Gaussian signals.
   }

\begin{document} 

%\maketitle  IS IGNORED %%%%%%%%%%%

%%%%%%%%%%%%%%%%%%%%%%%%%%%%%%%%%%%%%%%%%%%%%%%%%%%%%%%%%%%%%%%%%%%%%%%%
%======================================================================%
%============================== INTRO =================================%
%======================================================================%
%%%%%%%%%%%%%%%%%%%%%%%%%%%%%%%%%%%%%%%%%%%%%%%%%%%%%%%%%%%%%%%%%%%%%%%%

\section{Introduction and Summary}

In the Effective Field Theory of Inflation (EFTofI) we assume that inflation is a period where time translation are spontaneously broken~\cite{Cheung:2007st}. There is therefore a mode, the Goldstone mode, $\pi$, associated with time-translations, which is light during inflation. This allows us to write a Lagrangian for the fluctuations without assuming any knowledge of the mechanism that drove this spontaneous symmetry breaking. Such a description is very useful because the statistics of the fluctuations represents the only observable we are actually testing related to the period of inflation. Only the spatial FRW curvature is an observable associated with the presence of a background quasi de Sitter epoch in our past, but we have so far only an upper bound to it. Furthermore,  not having to assume anything of the mechanism driving the spontaneous symmetry breaking can be very important because the description of the actual mechanism might be beyond our analytical capabilities. The fact that inflation describes a theory around a vacuum that breaks time diffeomorphisms (diffs), while the same theory has a different vacuum that preserves the full diffs, implies that if we try to describe the theory of inflation starting from the theory around the Lorentz invariant vacuum, we have to be able to trust the theory as we traverse a possibly long path in the moduli space of the theory. It is not guaranteed at all that even though at both ends of the path the theory is weakly coupled, it is so along the path. We have drawn many lessons of how this can happen from dualities in supersymmetric field theories. Therefore, having a description directly for the inflationary perturbations is very useful.

While if we assume that inflation is a period were time translations are spontaneously broken we are guaranteed to have a light mode $\pi$, this does not mean that this must be the only light field in the theory. To have additional light scalar fields in a theory that is not tuned, the additional scalars need to be protected by some symmetries. This was the spirit of the so-called Effective Field Theory of Multifield inflation~\cite{Senatore:2010wk}, where additional scalar fields were introduced and their lightness was protected by either assuming that they were Goldstone bosons of some global internal symmetry group being spontaneously broken, or by supersymmetry~\cite{Senatore:2010wk}. Being light, these fields generated scale invariant perturbations, which then could directly affect the curvature or the isocurvature perturbations after horizon crossing. In the so-called quasi-single field models~\cite{Chen:2009zp,Noumi:2012vr, Green:2013rd}, it was noticed that fields could produce observable effects not only by producing super-horizon scale invariant perturbations, but also by affecting the inflaton directly, for example by mixing with it. Such models produce interesting signatures in the so-called squeezed limit of the three-point function when the additional scalar fields have a mass comparable to the Hubble rate during inflation and have sizable self-interactions (though this set up is in general quite ad-hoc). In a similar context, one can also consider higher spin particles, as recently done in~\cite{Arkani-Hamed:2015bza}. 

In this paper we focus on a class of mechanisms that produces an additional scalar field with mass close to $\sqrt{2} H$  during inflation  in a natural way. This is achieved by exploiting a somewhat peculiar symmetry breaking pattern, which we now explain and that does not involve the breaking of global internal symmetries. In four spacetime dimensions, diffeomorphisms are realized by four spacetime functions $\xi^\mu(x)$ as $x^\mu\to x^\mu+\xi^\mu(x)$. This implies that all diffeomorphisms can be non-linearly realized by the introduction of four Goldstone bosons. However, if we focus on the global limit of this gauge symmetry, we have Lorentz and spacetime translations, suggesting the presence of a higher number of possible Goldstone bosons. In fact, this is confirmed by the fact that if we introduce the vierbeins, we have that Lorentz and translations are independently gauged, which suggests that we could have a larger number of Goldstone bosons, depending of the symmetry breaking pattern. The explanation of this apparent paradox lies in what is called `inverse Higgs mechanism'~(see for example~\cite{Ivanov:1975zq,InverseHiggs,McArthur:2010zm,Nicolis:2013sga,Hidaka:2014fra}), which simply states that when we spontaneously break for example all spacetime symmetries, we must have at least four Goldstone bosons, while the additional Goldstone bosons can have a mass. Similar considerations apply to the case where one breaks only a fraction of the spacetime symmetries.

The situation can be most easily explained with an example (see~\cite{Donnelly:2010cr,Solomon:2013iza} for some earlier applications to inflationary cosmology). Imagine that we have a Lorentz invariant theory of a vector boson $\tilde A_\mu=e_\mu{}^a A_a$, where $e_\mu{}^a$ is the vierbein fields that we have introduced for convenience. $A_a$ is a scalar under spacetime diffs, and it is a vector under local boosts. Let us now imagine that $A_{\bar0}$ takes a constant vev ($\bar0$ is the time component of the local Lorentz index). This configuration will break local boosts, but will not break time diffs.

The three Goldstone bosons associated with this configuration, that we call $\eta^i$, can be considered as the ones associated with the local fluctuations of the direction in which the vev of $A_a$ is pointing. Let us imagine now to add to this theory a scalar field $\phi$ rolling down its potential with a time-dependent vev. This configuration now breaks time-diffs, so that we actually have an additional Goldstone boson, $\pi$, non-linearly realizing them. The field~$\pi$ is the usual Goldstone boson present in the Effective Field Theory of Inflation. Notice that the rolling scalar field breaks boosts as well, so that the field $\pi$ non-linearly realizes boosts, and not just time-diffs, on $\phi$. This suggests indeed that since $\pi$ can non-linearly realize boosts, the symmetry breaking pattern in which we break both boosts and time translations does not force the $\eta^i$ to be massless, and indeed, if the $\eta^i$ are very heavy with respect to the scales of interest, they can actually be integrated out. In other words,  in the case of spacetime symmetries, the symmetry breaking pattern does not completely determine the spectrum of Goldstone bosons.

The purpose of this paper is to study the just-described symmetry breaking pattern in the implementation where the breaking of boosts and time-translations requires additional Goldstone bosons $\eta^i$ beyond $\pi$, in the context of the Effective Field Theory of Multifield Inflation. While~\cite{Donnelly:2010cr,Solomon:2013iza} studied a particular implementation of this symmetry breaking pattern in inflationary cosmology at the level of the linear fluctuations, we here study the general EFT and focus on the non-Gaussian signatures.  
 We will find that the Goldstone boson $\pi$ will be coupled to the additional fields~$\eta^i$ in a rather unusual way. In fact, due to the peculiar transformation properties under diffs of $\pi$ and~$\eta^i$, $\pi$ can mix with the longitudinal component of the $\eta^i=\d^i\tilde\sigma$ fields, which is dynamical~\footnote{This is to be distinguished from what happens in Lorentz invariant theories of massive gauge bosons, where the component $\partial_\mu A^\mu$ is not dynamical.}. Furthermore, in the limit in which this mixing is weak, the $\eta_i$ have a mass that is approximately equal to $\sqrt{2}H$, which arises from their coupling to the spacetime curvature. This is a technically natural value in the vast majority of the parameter space. To our knowledge, this offers one of the very few ways to obtain fields with sizable interactions and with mass of order $H$ during inflation. Because of their mass,  the fields $\eta^i$ will not acquire scale invariant fluctuations, which implies that they can be detected only through their effect on~$\pi$. This is possible thanks to their mixing with~$\pi$ and to their self-interactions, that will lead to interesting, and potentially detectable, non-Gaussian signals.

%%%%%%%%%%%%%%%%%%%%%%%%%%%%%%%%%%%%%%%%%%%%%%%%%%%%%%%%%%%%%%%%%%%%%%%%
%======================================================================%
%============================== MODEL =================================%
%======================================================================%
%%%%%%%%%%%%%%%%%%%%%%%%%%%%%%%%%%%%%%%%%%%%%%%%%%%%%%%%%%%%%%%%%%%%%%%%

\section{EFT Construction} 

%======================================================================%
%====================== Unitary gauge action ==========================%
%======================================================================%
\subsection{Unitary gauge action}

For the purpose of describing the symmetry breaking pattern associated with the breaking of both time diffs and Lorentz boosts, it is useful to introduce the vierbein, defined as
\begin{equation}
g_{\mu\nu}= e_\mu^a e_{\nu}^b \eta_{ab}\ ,
\end{equation}
where $g_{\mu\nu}$ is the spacetime metric and $\eta_{ab}$ is the Minkowski one. The vierbein has more independent components than the metric, but this construction has six local Lorentz transformations that keep the spacetime metric unchanged. These eat the additional degrees of freedom contained in the vierbein, thus maintaining the original overall number of degrees of freedom. However, the introduction of the veilbein allows us to define local Lorentz transformations that set to zero the fluctuations is some fields that break boosts, without breaking time-translations. As we discussed in the introduction, the purpose of our paper is to include to the EFTofI treatment the case where, on top of the usual inflaton $\pi$, there is an additional matter sector with this property.

Once fluctuations around the symmetry breaking order parameter are removed with a spacetime dependent time diffeomorphism and a local boost, the remaining modes are all contained in the vierbein. The resulting action is only invariant under the residual symmetries, local rotations and time-dependent spatial diffeomorphisms, and has the form
\begin{equation}\label{S}
S = \int d^4x \sqrt{-g} \, \mathcal L (e_\mu^a,\nabla_\mu, \delta_\mu^0, \delta^{\bar 0}_a; t) \, .
\end{equation}
Here $\mu,\nu = 0,..,3$ and $a,b=\bar 0 , .. , \bar 3$ denote respectively diff and local Lorentz indices. We will be interested in the perturbations around an FRW background $\bar e_\mu^a(t)={\rm diag}(1,a,a,a)$. Eq.~\eqref{S} can be separated into pieces that break different symmetries
\begin{equation}
S=S_{EH} + S_P + S_L + S_{PL}\ ,
\end{equation}
where $S_{EH}$ is the Einstein-Hilbert action, $S_P$ contains only time diff breaking terms, $S_L$ contains only local Boost breaking terms and $S_{PL}$ contains terms that break both symmetries. As we will see shortly, in the decoupling limit the last three terms contain respectively the Nambu-Goldstone modes of time translation ($\pi$), those of local boosts ($\eta^i$) and mixing terms ($\pi \eta$).  

\paragraph{Time diff breaking sector:} The study of $S_P$ is done in the original EFTofI \cite{Cheung:2007st} to which we refer the reader for more details. Since this sector is invariant under local Lorentz, it can be expressed without use of the vierbein. Up to quadratic order in perturbations it is given by 
\begin{equation}\label{S_P}
S_{EH}+S_{P}=\int d^4x\sqrt{-g}\left[
	\frac{M_{\rm Pl}^2}{2}R
	-c(t)g^{00}
	-\Lambda(t)
	+\frac{M_2(t)^4}{2}(\delta g^{00})^2
+\ldots \right]\,,
\end{equation}
where $\delta g^{00}=g^{00}+1$ and we neglected higher derivative terms which can be expressed in terms of the extrinsic curvature $K_{\mu\nu}$ of the time slices
\begin{equation}
\delta g^{00}\delta K,\quad  \delta K^2,\quad (\delta K_{\mu\nu})^2,\quad  g^{\mu\nu}\d_\mu g^{00}\d_\nu g^{00},\quad R^{00}, \, \ldots \ .
\end{equation}
These operators can have important contributions in the de Sitter limit $\dot H\to 0$ where the Nambu-Goldstone mode of time diffs has a non-relativistic dispersion relation $\omega\sim k^2$ or $\omega\sim c_s k$ with $c_s\ll 1$~\cite{Cheung:2007st}. Though the formalism is general, for the purpose of this paper, we will not consider this limit and assume that the scalar Nambu-Goldstone mode has a dispersion with $c_s \simeq 1$, which gives the simple power counting rule $\d_t \sim \d_i \sim E$.

\paragraph{Boost breaking sector:} $S_L$ must contain all diff invariant terms built from the vierbein~$e_\mu^a$, the covariant derivative~$\nabla_\mu$ and the local Lorentz vector $\delta^{\bar 0}_a$.  Up to second order in derivatives the most general action is then
\begin{equation}\label{eq:action0}
S_{L}=\int d^4x\sqrt{-g}\left[
	-\frac{\alpha_1}{4}\left(\partial_\mu e_\nu^{\bar0}-\partial_\nu e_\mu^{\bar0}\right)^2
	-\frac{\alpha_2}{2} \left(\nabla^\mu e_\mu^{\bar0}\right)^2
	-\frac{\alpha_3}{2}\left(\nabla_\mu e_\nu^{\bar0}\right)^2
	-\frac{\alpha_4}{2}\left(e^\nu_{\bar0}\nabla_\nu e_\mu^{\bar0}\right)^2 + \ldots
\right]\,.
\end{equation}
Because we are using the full vierbein, this action contains background terms as well as terms that are only linear in fluctuations. These will affect the background equations of motion: below we will solve them such that all tadpoles disappear around the desired FRW background~\footnote{One could alternatively have chosen to parametrize the boost breaking sector in terms of $\delta e_\mu^{\bar 0} = e_\mu^{\bar 0} + n_\mu$, that starts linearly in the fluctuations. However this happens at the cost of introducing the vector $n_\mu$ in the boost-breaking action, which also breaks time-diffs, making the two symmetry breaking sectors less transparent. Another option is to define a fluctuating derivative of the vierbein as
\begin{align}\nonumber
D_{\mu\nu}=\nabla_\mu e_\nu^{\bar0}+H\left(g_{\mu\nu}+e_\mu^{\bar0}e_\nu^{\bar0}\right)\,,
\end{align}
which vanishes on the background and is covariant under the full diffs (apart from the time dependence of $H$). This would simplify a bit the power counting of the fluctuations and would allow us to eliminate the contribution to the FRW tadpoles. However, we find that the parametrization in (\ref{eq:action0}) is simple enough for the purposes of this paper.
}.

Since the full unitary gauge action is not invariant under time diffs, the coefficients $\alpha_i$ can in general be time dependent. Although these terms now break all symmetries and should therefore be included in $S_{PL}$, we will keep them in $S_L$ for simplicity. It is technically natural to take the time-dependence of these coefficients, as well as of all the other coefficients in the full Lagrangian, to be small, of order of the slow roll parameters $\dot\alpha/(H\alpha)\sim \epsilon$, and in practice negligible, which is what we will do in this paper~\footnote{See~\cite{Behbahani:2011it} for a study of the case where the time dependence of the EFTofI parameters is non-negligible.}. Here and for the rest of the paper we take all the $\alpha_i$'s to be of the same order, generically referred to as $\alpha$.

If one is interested in a theory where time diffs are not spontaneously broken, then the theory is described by the action $S_{EH}+S_L$ with the $\alpha_i$ constant. This model describes ``framids'' (recently discussed in \cite{Nicolis:2015sra}) which here are coupled to gravity. This action can also be established with a coset construction, as is shown in appendix \ref{app:coset}.

\paragraph{Mixing sector:} We finally turn to terms that break both time diffs and local boosts. Writing $\delta e_\mu^{\bar 0} = e_\mu^{\bar 0} + n_\mu$, where $n_\mu$ is the unit vector perpendicular to the time slices (in unitary gauge $n_\mu = - \delta_\mu^0/\sqrt{-g^{00}}$), we have
\begin{equation}\label{S_PL}
S_{PL}=\int d^4x\sqrt{-g}\,\left[
	\frac{-\beta_1}{2}\left(\delta e_\mu^{\bar0}\right)^2 
	-\frac{\beta_2}{2}\delta g^{00}\left(\nabla^\mu e_\mu^{\bar 0}+3H\right)+\ldots
\right]\,,
\end{equation}
where, as discussed, the coefficients $\beta_i(t)$ can again depend on time. The term proportional to Hubble in $\beta_2$ was added for simplicity to remove the background $\left.\nabla^\mu n_\mu\right|_0=3H$.

\paragraph{Tadpole Terms:} The coefficients of the terms that start at linear order in fluctuations are the only ones that contribute to the background equations of motion; they can be fixed by requiring that FRW be a solution to $\delta(S_{EH} + S_P + S_L + S_{PL})=0$. This imposes 
\begin{equation}\label{background}
\frac{1}{2}\left(3Hf+\dot{f}\right)-\Lambda=-M_{\rm Pl}^2(3H^2+\dot{H})\,,
\quad
\frac{1}{2}\dot{f} +c=-M_{\rm Pl}^2\dot{H}
\quad
{\rm with}
\quad
f=3H\alpha_2+H\alpha_3\,.
\end{equation}
We choose to keep $\alpha_2,\,\alpha_3$ unconstrained and solve \eqref{background} for $\Lambda$ and $c$. Using $\dot\alpha/\alpha\sim \epsilon H$, the solution is then of the form
\begin{equation}\label{cLambda}
c = \epsilon H^2 M_{\rm Pl}^2 \left(1+\O(\alpha/M_{Pl}^2)\right) \, , \quad
\Lambda =  H^2M_{\rm Pl}^2(3 -\epsilon ) \left(1+\O(\alpha/M_{Pl}^2)\right) \, .
\end{equation}
We will be interested in theories where the boost breaking sector becomes strongly coupled much before the Planck scale: $\alpha \ll M_{\rm Pl}^2$. In such cases, the background solution \eqref{cLambda} reduces to that of the EFTofI \cite{Cheung:2007st}. 

%======================================================================%
%==================== Introducing the Nambu-Goldstone fields =======================%
%======================================================================%
\subsection{Introducing the Nambu-Goldstone fields}
Full, non-linearly realized invariance under the original gauge symmetries can be recovered from the unitary gauge action with the St\"uckelberg trick. This is done by performing a local broken transformation, promoting the transformation parameters to fields, and realizing the broken symmetries nonlinearly on these fields. For time diffeomorphisms, this amounts to applying the transformation
\begin{equation}
x^\mu\rightarrow x'{}^{\mu}(x) = (t + \pi(x'),x^i)\ ,
\end{equation}
and postulating that $\pi$ transforms non-linearly
\begin{subequations}
\begin{align}
x^\mu	&\rightarrow x'{}^{\mu}(x) = x^\mu + \xi^\mu(x)\, , \\
\pi(x)	&\rightarrow \pi'(x') = \pi(x) - \xi^0(x)\, .
\end{align}
\end{subequations}
To restore invariance under local boosts, we act on local Lorentz indices with an element of the quotient $SO(1,3)/SO(3)$ which can be parametrized as
\begin{equation}
\Lambda^a{}_b (x) \equiv \left(e^{\eta^i K_i}\right)^a{}_b\, ,
\end{equation}
where $K_i$ are the boost generators in the 4-vector representation. Invariance is recovered if the $\eta^i$ transform nonlinearly under a local Lorentz transformation $ g$
\begin{equation}
e^{\eta(x)\cdot K} \rightarrow e^{\eta'(x')\cdot K} =h(g,\eta)   e^{\eta(x)\cdot K} g^{-1}\, ,
\end{equation}
where $h(g,\eta)$ is a Nambu-Goldstone dependent local \emph{rotation} that ensures that the RHS is still a boost~\footnote{If $g$ is a rotation then $h(g,\eta)=g$ and the $\eta^i$ transform {\em linearly}, under the spin 1 representation.}. This implies in particular that $\Lambda^{\bar 0}{}_a e_\mu^{a}\equiv Q_\mu$ is a scalar under local Lorentz (and a vector under diffs). Notice that, as typical for Goldstone bosons, the way the $\eta$ and $\pi$ fields transform is independent of the representation under which the order parameters transform. This is yet another advantage of the EFT formalism~\footnote{Notice that if we keep the metric dynamical, $\eta^i$'s are three scalar fields under diffs. However, when we later focus on the decoupling limit and fix the background metric, the $\eta^i$ will inherit the transformation under rotation of the metric, and will transform as 3-vectors.}. 

Both of these transformations amount to replacing
\begin{align}\label{Q_def}
t 	\	&\to\  t+ \pi\ , \nonumber \\
g^{00} \ 	&\to \ g^{\mu\nu}\d_\mu(t+ \pi)\d_\nu (t+ \pi)\equiv Q \ ,\\ \nonumber 
e_\mu^{\bar 0} \
		&\to\  \Lambda^{\bar 0}{}_a e_\mu^{a} \equiv Q_\mu =  (\delta^{\bar 0}_a + \eta_i\delta^{i}_a + \frac{1}{2}\eta^2\delta^{\bar 0}_a + \O(\eta^3))e_\mu^{a} \ ,
\end{align}
in the action, which gives
\begin{subequations}\label{S_stu}
\begin{align}
S&=S_{EH}+S_{P}+S_{L}+S_{PL}\,,
\\
\label{S_P}
S_{P}&=\int d^4x\sqrt{-g}\left[
	-c \, Q -\Lambda+\frac{M_2^4}{2}(Q+1)^2+\ldots
\right]\,,
\\
\label{S_L}
S_{L}&=\int d^4x\sqrt{-g}
\left[
-\frac{\alpha_1}{4}\left(\partial_\mu Q_\nu -\partial_\nu Q_\mu\right)^2
-\frac{\alpha_2}{2} \left(\nabla^\mu Q_\mu\right)^2\right. \\ \nonumber
& \hspace{75pt} \left. -\frac{\alpha_3}{2}\left(\nabla_\mu Q_\nu\right)^2
-\frac{\alpha_4}{2}\left(Q^\nu\nabla_\nu Q_\mu\right)^2 + \dots
\right]\,,
\\
\label{S_PL}
S_{PL}&%=\int d^4x\sqrt{-g}\,(-\beta_1)\left(e_\mu^{\bar0}n^\mu-1\right)
=\int d^4x\sqrt{-g}\,\left[
\beta_1\left(\frac{Q^\mu\d_\mu(t+\pi)}{\sqrt{-Q}}+1 \right)
%2\beta_1\left({Q^\mu\d_\mu(t+\pi)}/Q -1 \right)
-\frac{\beta_2}{2}(Q+1)\left(\nabla^\mu Q_\mu+3H\right)+\ldots \right]\,,
\end{align}
\end{subequations}
where all the time dependent coefficients ($c,\Lambda,M_2,\alpha_i,\beta_i$) are evaluated at $t+\pi$. The action is now manifestly invariant under all diffs and local Lorentz, and contains four St\"uckelberg fields on top of the metric: the time diff Nambu-Goldstone mode $\pi$ and the boost Nambu-Goldstone modes $\eta_i$.

%\paragraph{Wess-Zumino term}
%A co-dimension 1 brane admits a single Wess-Zumino term which corresponds the volume {\em enclosed} by the brane.\marginpar{\LD{$\bigstar$}} In terms of the Nambu-Goldstone field and in the decoupling limit, it is expressed as $S_{WZ} = \int d^4 x \, \pi$ . From symmetry principles, such a term should also be added to the present effective theory -- however since it is a tadpole it must be set to zero for FRW to be a solution. 

%======================================================================%
%================= Action in the decoupling limit =====================%
%======================================================================%
\subsection{Action in the Decoupling Limit}

Notice that the quadratic action \eqref{S_stu} contains mixing terms between the Nambu-Goldstone fields and the metric. Since some components of the metric are pure gauge modes or constrained variables, the dynamics resulting from the action is not completely transparent yet. In this section, we determine the regime where gravitational fluctuations can be ignored, and find the decoupled action for the Nambu-Goldstone modes. The result \eqref{S_DL} below will be the starting point for the study of stability and non-Gaussianities in the following sections.

At second order in the perturbations, by rotational invariance the Nambu-Goldstone fields can only couple to the spin 0 and spin 1 modes of the metric (to see this, integrate spatial derivatives by parts until they all act on the metric). Of the 4 scalar modes of the metric $N, \d_i N^i, h^i_i$ and $\d_i\d_j\delta h_{ij}$ (in ADM parametrization), two can be removed with transformations $t\to t+ \xi^0$ and $x_i \to x_i + \d_i \xi$, and the remaining two are constrained variables. Similarly, the metric contains two spin 1 modes $N_i^\perp$ and $\d_i h_{ij}$, one of which can be removed with the transformation $x^i\to x^i + \xi^i$, the other being a constrained variable. The correct way to deal with the mixing with the non-dynamical components is to solve their constraint equations, and then insert back into the action -- this is done in Appendix~\ref{app:ADM}.   The result is a quadratic action directly for the spin 2 part of the metric and the Nambu-Goldstone modes. One can then consistently choose to study the Nambu-Goldstone sector on a fixed background without exciting metric fluctuations. The main result of Appendix \ref{app:ADM} is that there exists a decoupling regime, as in spontaneously broken gauge theories, in which metric fluctuations can be ignored from the start. The first condition is that the typicial energies be larger than the mixing energy 
\begin{equation}\label{DL1}
E\gg E_{\rm mix} \sim \epsilon ^{1/2} H\, , 
\end{equation}
as was found in the single field EFTofI \cite{Cheung:2007st,Cheung:2007sv}. The additional condition when local boosts are spontaneously broken is that the coefficients in the action satisfy
\begin{equation}\label{DL2}
\alpha \ll M_{\rm Pl}^2 \, ,\quad \sqrt{\epsilon} \sqrt{\alpha}/M_{\rm Pl} \ll \beta_2^c \ll \min(1/\sqrt{\epsilon}, M_{\rm Pl}/\sqrt{\alpha})\ .
\end{equation}
The canonical mixing parameter $\beta_2^c \sim \beta_2/(\sqrt{\epsilon\alpha} H M_{\rm Pl})$ is defined precisely later in \eqref{beta_canon}, all that is needed here is to know that it is a dimensionless coefficient that characterizes $\eta\pi$ mixing. The first condition in \eqref{DL2} can be simply thought of as requiring that the strong coupling scale in the boost breaking sector be lower than the Planck scale. The second one forbids too strong mixing on one hand (although it still allows $\beta_2^c \gg 1$), and on the other it requires the original $\eta\pi$ mixing to be at least as large as the one induced by gravity mixing which is of order $\sqrt{\epsilon} \sqrt{\alpha}/M_{\rm Pl}$ (this is so because for phenomenological reasons, we will be interested in having a non-negligible mixing begtween $\pi$ and $\eta$, and so we ensure that the leading mixing is not the one induced by gravity). In the slow-roll inflation limit none of these conditions are restrictive, so for the rest of the analysis we will focus on regimes where \eqref{DL1} and \eqref{DL2} are satisfied and leave the exploration of other regimes for future work.

The action in the decoupling limit can easily be obtained from \eqref{S_stu} by fixing the metric to the FRW background $e_\mu^a={\rm diag}(1,a,a,a)$, i.e. with the replacement
\begin{subequations}
\begin{align}
Q		&\to -1 -2\dot \pi - \dot \pi^2 + (\d_i \pi/a)^2 \\
Q_\mu	&\to \delta_\mu^0(1+\eta^2/2) + \delta_\mu^i(a\eta_i) + \O(\eta^3) \, ,
\end{align}
\end{subequations}
and similarly for the higher derivative operators.
Up to quadratic order in the Nambu-Goldstone fields, this leads to the following action \footnote{
Because the combination $\delta e_\mu^0= \delta_\mu^i(\eta_i-\d_i \pi) + \ldots$ linearly realizes all symmetries, it is possible to define $\tilde\eta^i=\eta^i-\d^i\pi$, which transform as a matter field, and use $\tilde\eta^i$ directly. The action (\ref{S_DL}), or more generally (\ref{S_stu}), can indeed be alternatively constructed by adding a spin-1 field to the EFTofI with the constraint $n^\mu A_\mu=0$. This can be implemented at the linear level with a Lagrangian of the form
\begin{eqnarray*}
&&\mathcal L \sim \alpha_1 (\d_{[\mu}A_{\nu]})^2 + \alpha_2 (\d_\mu A^\mu)^2 + \alpha_3 (\d_\mu A_\nu)^2 + \alpha_4 (\d^0 A_\mu)^2 + m^2 A_\mu^2 + M^2(A^0)^2\, \\ \nonumber
&&\qquad + \beta_1(\d_\mu\pi + A_\mu)^2 + \beta_2 \d^0 \pi (\d_\mu A^\mu)\ ,
\end{eqnarray*}
with $M\to \infty$. 

It is in the sense of this peculiar limit of the parameter space of the Lagrangians where we couple the Goldstone boson $\pi$ to an additional vector field that our construction in terms of Goldstone bosons of Boosts should be understood.   Indeed, it is interesting to compare this to the theory studied in Ref.~\cite{Arkani-Hamed:2015bza}, where they take the opposite limit, $M=0$, and furthermore take $m\sim H$. There, the authors impose the constraint $\d_\mu A^\mu=0$, which makes the linear mixing between $\pi$ and $\tilde \eta_i\sim A_i$ vanish.  It would be interesting to study the more general consistent theories with $M\neq 0$ and $\d_\mu A^\mu\neq 0$. We leave this to future work. We thank   Pietro Baratella and Paolo Creminelli for discussions about this point.
}
\begin{subequations}\label{S_DL}
\begin{align}
%\label{S_P_NG}
S_{P}&=\int d^4x\,a^3\bigg[
\left(c+2M_2^4\right)\dot{\pi}^2-c\frac{(\partial_i\pi)^2}{a^2}+\dots
\bigg]\,,
\\*\label{S_L_NG}
S_{L}&=\\ \nonumber
\int d^4x&\,a^3
\left[
\frac{\alpha_1+\alpha_3-\alpha_4}{2}\left(\dot{\eta}_i^2-2H^2\eta^2_i\right)
-\frac{\alpha_1}{4}\frac{\left(\partial_i\eta_j-\partial_j\eta_i\right)^2}{a^2}
-\frac{\alpha_2}{2} \frac{(\partial_i\eta_i)^2}{a^2}
-\frac{\alpha_3}{2}\frac{(\partial_i\eta_j)^2}{a^2}+\ldots
\right],
\\
\label{S_PL_NG}
S_{PL}&=\int d^4x\,a^3 \left[\frac{-\beta_1}{2}\left(\eta_i-\frac{\partial_i\pi}{a}\right)^2 + \beta_2\dot \pi \frac{\partial_i\eta^i}{a}+\dots\right]
\,,
\end{align}
\end{subequations}
where we dropped subleading terms in the slow-roll expansion. Notice that $S_{PL}$ contains a mass term for the boost Nambu-Goldstone mode $\eta$ which does not vanish in the flat space limit. It is indeed a generic feature of spontaneous breaking of spacetime symmetries that some of the associated Nambu-Goldstone modes can be gapped \cite{Low:2001bw}. Although $\eta_i$ transforms non-linearly under boosts, so does $\d_i \pi$; this allows one to write a mass term for $\eta$ and to let the boosts be non-linearly implemented just by $\pi$. Notice that, obviously, this mass term does not exist if only boosts are broken as, without $\eta^i$'s, there would be no other field available to non-linearly realize them.

%======================================================================%
%=============== Large $\eta$ mass regime and EFTofI ===================%
%======================================================================%
\subsection{Large $\eta$ mass regime and EFTofI\label{sec:large_mass}}

We found in the previous sections that the boost Nambu-Goldstone bosons $\eta^i$ have a mass
\begin{equation}
m_{\eta}^2 = 2H^2 + \frac{\beta_1}{\alpha_1 + \alpha_3 - \alpha_4}\, .
\end{equation}
If $\beta_1\gg H^2 \alpha$, then at energies of order Hubble or lower, $\eta$ can be integrated out in the action and one obtains an effective action for $\pi$. Since we are not changing the background in this process, we expect to recover the usual EFTofI with coefficients determined from the initial action. In this section, we explain exactly how this happens and find these coefficients.

From the unitary gauge action \eqref{S_PL} or the St\"uckelbergized one
\begin{equation}
S_{PL} \ni \int d^4 x \, a^3 \left[\frac{-\beta_1}{2}(e_\mu^{\bar 0}+n_\mu)^2\right]
	\quad\xrightarrow{\hbox{\rm St\"u }}\quad \int d^4 x  \, a^3 \left[\frac{-\beta_1}{2}(\eta_i - \d_i\pi/a +\ldots)^2\right] \ ,
\end{equation}
we can see that at energies $E^2\ll \beta_1/\alpha$ this term dominates the $\eta$ kinetic term and we can simply integrate out the boost mode with the following replacement in the action
\begin{equation}
e_\mu^{\bar 0} \to -n_\mu\,  \quad \hbox{or} \quad
\eta_i \to \d_i\pi/a +\ldots\, .
\end{equation}
Here the dots stand for higher order terms in $\pi$ and $\eta$. This expression might be recognized as an ``inverse-Higgs constraint'' in the context of spontaneously broken spacetime symmetries \cite{Ivanov:1975zq}. The various terms in the unitary action now become 
\begin{subequations}
\begin{align}
&\alpha_1 : \quad (\d_{[\nu}e_{\mu]}^{\bar 0})^2 
	\to (\d_{[\nu}n_{\mu]})^2  = \frac{1}{8} (g^{00})^{-2}h^{\mu\nu}\d_\mu g^{00}\d_\nu g^{00}\\
&\alpha_2 : \quad (\nabla^\mu e_\mu^{\bar 0})^2
	\to (\nabla^\mu n_\mu)^2  = K^2\\
&\alpha_3 : \quad (\nabla_\mu e_\nu^{\bar 0})^2
	\to (\nabla_\mu n_\nu)^2  = (K_{\mu\nu})^2 -  \frac{1}{4}(g^{00})^{-2}h^{\mu\nu}\d_\mu g^{00}\d_\nu g^{00} \\
&\alpha_4 : \quad (e_{\bar 0}^\mu\nabla_\mu e_\nu^{\bar 0})^2
	\to (n^\mu\nabla_\mu n_\nu)^2  = \frac{1}{4}(g^{00})^{-2}h^{\mu\nu}\d_\mu g^{00}\d_\nu g^{00} \\
&\beta_2 : \quad \delta g^{00} \nabla^\mu e_\mu^{\bar 0}
	\to \delta g^{00} \nabla^\mu n_\mu  = \delta g^{00} K 
\end{align}
\end{subequations}
Notice that only four different terms are generated and are given by
\begin{equation}\label{equ:EFTofI_higherderiv}
\delta g^{00} K,\quad   K^2,\quad (K_{\mu\nu})^2,\quad  (g^{00})^{-2}h^{\mu\nu}\d_\mu g^{00}\d_\nu g^{00}\, ,\
\end{equation}
this is exactly what one would expect from the single field EFTofI. Those operators generate the EFTofI coefficients, $\bar M_1^3 = \beta_2$, $\bar M_2^2 = \alpha_2$, $\bar M_3^2 = \alpha_3$ and $\bar M_4^2 = \alpha_1/4 - \alpha_3 + \alpha_4$ (here $\bar M_4^2$ is the coefficient of the last term in \eqref{equ:EFTofI_higherderiv}) as well as modifying the coefficients of $(\delta g^{00})^n$ ($n\geq2$).
%

%%%%%%%%%%%%%%%%%%%%%%%%%%%%%%%%%%%%%%%%%%%%%%%%%%%%%%%%%%%%%%%%%%%%%%%%
%======================================================================%
%=============== CONSTRAINTS AND NON-GAUSSIANITIES ====================%
%======================================================================%
%%%%%%%%%%%%%%%%%%%%%%%%%%%%%%%%%%%%%%%%%%%%%%%%%%%%%%%%%%%%%%%%%%%%%%%%

\section{Stability and Superluminality Constraints}

In the previous section
we introduced the generic effective action
for our symmetry breaking pattern.
Our construction was based on the symmetry structure only
and contains free parameters (free functions of time to be precise),
which should be further constrained
by physical requirements
such as stability of the background.

In flat space, superluminalities in EFTs are known to obstruct Lorentz invariant UV completions \cite{Adams:2006sv}. The situation is more subtle in curved space, where field redefinitions called disformal transformations
%-- field dependent change of coordinates --
change propagation speeds. However ratios of speeds are preserved under these transformations: one reasonable condition to impose is thus that all fields propagate at most as fast as the metric tensor modes: $c_s \leq c_\gamma$. As we will see, in the decoupling limit we have $c_\gamma = 1$, so this corresponds simply to imposing subluminality in the naive sense.

In this section
we take a closer look at the second order action
and clarify
physically reasonable parameter regimes
of the model (see also~\cite{Donnelly:2010cr,Solomon:2013iza}).
Throughout this section
we use the action in the decoupling limit.

\subsection{Spin 1 and 2 sector}

As a warmup, let us begin with the second order action
for the spin 1 and 2 sector. The dynamical degrees of freedom
in our setup
are the Nambu-Goldstone modes, $\pi$ and $\eta_i$,
for  time diffs and local boosts,
and the gravitational tensor mode, $\gamma_{ij}$.
$\eta_i$ can be further decomposed into a spin $1$ component, $\eta_{\bot i}$,
and a spin $0$ component, $\eta_{\parallel i}=\partial_i\tilde{\sigma}$
as 
\begin{align}
\eta_i=\eta_{\bot i}+\partial_i\tilde{\sigma}
\quad
{\rm with}
\quad
\partial_i\eta_{\bot i}=0\,.
\end{align}
In momentum space,
the second order action for the spin $1$ mode, $\eta_{\bot i}$,
and the spin $2$ mode, $\gamma_{ij}$,
is simply given by
\begin{align}
S_1&=\int d\tau\int \frac{d^3{\bf k}}{(2\pi)^3}\,a^2
\frac{\bar{M}^2}{2}\left[
\eta'_{\bot i{\bf \,k}}\eta'_{\bot i{\bf \,-k}}
-\left(c_{\eta_\bot}^2k^2+a^2m_\eta^2\right)\eta_{\bot i{\bf \,k}}\eta_{\bot i{\bf \,-k}}
\right]\,,
\\
S_2&=\int d\tau\int \frac{d^3{\bf k}}{(2\pi)^3}\,a^2
\frac{M_{\rm Pl}^2c_\gamma^{-2}}{8}\left[
\gamma'_{ij{\bf \,k}}\gamma'_{ij{\bf \,-k}}
-c_{\gamma}^2k^2\,\gamma_{ij{\bf \,k}}\gamma_{ij{\bf \,-k}}
\right]\,.
\end{align}
Here and in the rest of this section
we use conformal time, $\tau$,
defined by $dt=ad\tau$.
The parameters,
$\bar{M}$, $c_{\eta_\bot}$, and $m_\eta$
are the normalization factor, the sound speed, and the effective mass
of the spin $1$ mode:
\begin{align}
\label{eta_parameters}
\bar{M}^2=\alpha_1+\alpha_3-\alpha_4\,,
\quad
c_{\eta_\bot}^2=\frac{\alpha_1+\alpha_3}{\alpha_1+\alpha_3-\alpha_4}\,,
\quad
m_\eta^2=2H^2+\frac{\beta_1}{\alpha_1+\alpha_3-\alpha_4}\,.
\end{align}
The sound speed $c_\gamma$ of the tensor mode can be read off the quadratic action \eqref{S_ADM} and is given by
\begin{align}
c_\gamma^2=\frac{M^2_{\rm Pl}}{M_{\rm Pl}^2-\alpha_3}\, .
\end{align}
Notice that $c_\gamma\simeq 1$ in the decoupling regime.
Passing to the $\eta^i$'s, we first require that
the temporal and spatial kinetic terms
have the correct sign.
It guarantees absence of ghosts
and the stability at subhorizon scales.
We also require that $\eta_{\bot i}$ has the positive mass squared,
which prohibits a tachyonic instability and guarantees the stability
at the superhorizon scale.
These conditions can be stated as
\begin{align}
\bar{M}^2>0\,,
\quad
c_{\eta_\bot}^2>0\,,
\quad
m_\eta^2>0\,,
\end{align}
and can be rephrased as
\begin{align}
\label{stability_for_12}
\alpha_1+\alpha_3-\alpha_4>0\,,
\quad
\alpha_1+\alpha_3>0\,,
\quad
\beta_1>-2H^2(\alpha_1+\alpha_3-\alpha_4)\,.
\end{align}
We further impose subluminality, leading to the constraints
\begin{equation}
c_{\eta_{\bot}}^2\leq1\,,
\quad
\leftrightarrow
\quad  \alpha_4 \leq 0\, ,
\end{equation}
where we used the stability conditions~\eqref{stability_for_12}.

\subsection{Spin 0 sector}
\label{subsec:spin0}

We then perform a similar discussion for the scalar sector.
The second order action for the spin $0$ modes is given by
\begin{align}
\nonumber
S_0&=\frac{1}{2}\int d\tau\int \frac{d^3{\bf k}}{(2\pi)^3}\,a^2
\bigg[
M^4\left(\pi'_{\bf k}\pi'_{\bf -k}
-c_\pi^2 k^2\,\pi_{\bf k}\pi_{\bf -k}
\right)
+\bar{M}^2\Big(\sigma'_{\bf k}\sigma'_{\bf -k}
-\left(c_\sigma^2k^2+a^2m_\eta^2\right)\sigma_{\bf k}\sigma_{\bf -k}
\Big)
\\
\label{scalar_second}
&\qquad\qquad\qquad\qquad\qquad
+2\beta_1ak\, \pi_{\bf k}\sigma_{\bf -k}
-2\beta_2k\, \pi'_{\bf k}\sigma_{\bf -k}
\bigg]\,,
\end{align}
where we introduced $\sigma_{\bf k}=k\tilde{\sigma}_{\bf k}$.
The parameters, $M$, $c_\pi$, and $c_\sigma$, are defined by
\begin{align}
M^4=2c+4M_2^4\,,
\quad
c_\pi^2=\frac{2c+\beta_1}{2c+4M_2^4}\,,
\quad
c_\sigma^2=\frac{\alpha_2+\alpha_3}{\alpha_1+\alpha_3-\alpha_4}\,.
\end{align}
In contrast to the spin $1$ and $2$ sector,
the scalar sector accommodates a kinetic mixing
between $\pi$ and $\sigma$.
Such a mixing interaction
makes the derivation of the spectrum rather complicated.
In the following
we clarify the stability conditions at the sub and superhorizon scales,
and the superluminality constraint
taking into account the mixing interactions appropriately.

\paragraph{Stability on subhorizon scales:}

We start by discussing stability on subhorizon scales. Let us write the action as
\begin{align}
\nonumber
S_0&=\frac{1}{2}\int d\tau\int \frac{d^3{\bf k}}{(2\pi)^3}\,a^2
\bigg[
\left(\pi'_{c\,{\bf k}}\pi'_{c\,{\bf -k}}
-c_\pi^2 k^2\,\pi_{c\,{\bf k}}\pi_{c\,{\bf -k}}
\right)
+\Big(\sigma'_{c\,{\bf k}}\sigma'_{c\,{\bf -k}}
-\left(c_\sigma^2k^2+a^2m_\eta^2\right)\sigma_{c\,{\bf k}}\sigma_{c\,{\bf -k}}
\Big)
\\
&\qquad\qquad\qquad\qquad\qquad
+2H\beta^c_1ak\, \pi_{c\,{\bf k}}\sigma_{c\,{\bf -k}}
-2\beta^c_2k\, \pi'_{c\,{\bf k}}\sigma_{c\,{\bf -k}}
\bigg]\,,
\end{align}
where the canonical fields
$\pi_c$ and $\sigma_c$ were defined as 
\begin{align}
\pi_{c\,{\bf k}}=M^2\pi_{\bf k}\,,
\quad
\sigma_{c\,{\bf k}}=\bar{M}\sigma_{\bf k}\,.
\end{align}
and where we introduced dimensionless couplings, $\beta_1^c$ and $\beta_2^c$, as
\begin{align}\label{beta_canon}
\beta_1^c = \frac{\beta_1}{M^2\bar{M} H} \quad {\rm and} \quad
\beta_2^c = \frac{\beta_2}{M^2\bar{M} }\, .
\end{align}
In the regime,
\begin{align}
c_\pi\frac{k}{a}\,,\,\,c_\sigma\frac{k}{a}
\,\,\gg\,\,
H\,,\,\,\beta_1^cH\,,\,\,m_\eta\,,
\end{align}

$\pi_c$ and $\sigma_c$ behave like massless fields
and only the $\beta_2^c$ mixing coupling becomes relevant:
\begin{align}
\nonumber
S_0&\simeq\frac{1}{2}\int d\tau\int \frac{d^3{\bf k}}{(2\pi)^3}\,a^2
\Big[
\big(\pi'_{c\,{\bf k}}\pi'_{c\,{\bf -k}}
-c_\pi^2 k^2\,\pi_{c\,{\bf k}}\pi_{c\,{\bf -k}}\big)
+\big(\sigma'_{c\,{\bf k}}\sigma'_{c\,{\bf -k}}
-c_\sigma^2k^2\sigma_{c\,{\bf k}}\sigma_{c\,{\bf -k}}\big)
\\
&\qquad\qquad\qquad\qquad\qquad
-2\beta^c_2k\, \pi'_{c\,{\bf k}}\sigma_{c\,{\bf -k}}
\Big]\,.
\end{align}
Since curvature effects are negligible,
the mode function approximately takes the form~$\sim~e^{\pm i\omega \tau}$.
The on-shell condition can then be stated as
\begin{equation}
\text{det}\left( \begin{array}{cc}
\omega^2-c_\pi^2k^2 & i\beta_2^c\,k\omega \\
-i\beta_2^c\,k\omega & \omega^2-c_\sigma^2k^2  \end{array} \right)=0
\quad
\leftrightarrow
\quad
\omega^4-\left(c_\pi^2+c_\sigma^2+(\beta_2^c)^2\right)k^2\omega^2+c_\pi^2c_\sigma^2k^4=0
\,,
\end{equation}
whose solution is given by
\begin{align}
\label{sound_speed_mixing}
\omega_\pm^2=c_\pm^2k^2
\quad
{\rm with}
\quad
c_\pm^2=\frac{c_\pi^2+c_\sigma^2+(\beta_2^c)^2\pm\sqrt{(c_\pi^2-c_\sigma^2)^2+2(\beta_2^c)^2(c_\pi^2+c_\sigma^2)+(\beta_2^c)^4}}{2}\,.
\end{align}
Here it should be emphasized that
the dispersion relation is linear
for any parameter choice.\footnote{
In Appendix~\ref{app:Sta}
we illustrate dispersion relations
in various inflation models with mixing interactions,
such as multi-field inflation and quasi-single field inflation,
and compare them to our setup.} We immediately realize that one of the modes is superluminal for $\beta_2^c\geq 1$, so that we focus on the regime $\beta_2^c\lesssim 1$. Requiring the time-kinetic energy to be positive, implies that 
\begin{equation}
M^4>0\ , \quad \bar M^2>0\ .
\end{equation}
The positivity of the gradient energy then requires  $c_\pm^2>0$.
More concretely,
this can be stated as
\begin{align}
c_\pi^2c_\sigma^2>0\,,
\quad
c_\pi^2+c_\sigma^2+(\beta_2^c)^2>0
\,.
\end{align}
Note that these conditions can be rephrased
in the regime, $\beta_2^c\ll |c_\pi|,|c_\sigma|$,
as
\begin{align}\label{stability_condition_beta}
c_\pi^2>0\,,
\quad
c_\sigma^2>0
\quad
\leftrightarrow
\quad
\beta_1^c\leq \frac{4M_2^4}{M^2\bar{M}H}\ ,
%\frac{2c}{2c+4M_2^4}+\frac{\bar{M}H}{(2c+4M_2^4)^{1/2}}\beta_1^c\leq1\,,
\quad
\alpha_2+\alpha_3>0\,,
\end{align}
where we used the conditions~\eqref{stability_for_12}.

It is useful to give the expressions for the dispersion relation for  $c_\pi=c_\sigma=c_s$.
First,
in the weak mixing regime, $|\beta_2^c|\ll c_s$,
the sound speed~\eqref{sound_speed_mixing} is given by
\begin{align}
c_\pm^2\simeq c^2_s\left(1\pm\frac{\beta_2^c}{c_s}\right)\,,
\end{align}
so that the superluminality constraint
is given by
\begin{align}
|\beta_2^c|\leq c_s(1-c_s^2)\,.
\end{align}
Notice that for $c_s=1$, $\beta_2^c$ needs to vanish. However, for a small $\beta_2^c$ to be allowed, one needs only a small departure of $c_s$ from unity. For generic $c_\pi$ and $c_\sigma$, and $\beta_2^c\ll1$, $c_\pi$ and $c_\sigma$
can be interpreted as the sound speed of $\pi$ and $\sigma$,
respectively.

In the regime $|\beta_2^c|\gg c_s$,
the sound speed~\eqref{sound_speed_mixing} takes the form,
\begin{align}
c_+^2\simeq (\beta_2^c)^2\,,
\quad
c_-^2\simeq \frac{c_s^4}{(\beta_2^c)^2}\,,
\end{align}
so that, in order to avoid superluminality, $\beta_2^c$ can only approach unity from below.

\paragraph{Stability on superhorizon scales:} On superhorizon scales, the gradient terms become negligible and the mass terms become relevant. In order to avoid a slow, but still unpleasant, tachyonic instability, the only additional requirement is that 
\begin{equation}
m_\eta^2>0 \ .
\end{equation}

%%%%%%%%%%%%%%%%%%%%%%%%%%%%%%%%%%%%%%%%%%%%%%%%%%%%%%%%%%%%%%%%%%%%%%%%
%======================================================================%
%========================== REGIMES ===================================%
%======================================================================%
%%%%%%%%%%%%%%%%%%%%%%%%%%%%%%%%%%%%%%%%%%%%%%%%%%%%%%%%%%%%%%%%%%%%%%%%

%\section{Regimes}

%======================================================================%
%====================== non-Gaussianities =============================%
%======================================================================%
\section{Non-Gaussianities}
\label{subsec:nonG}
%\LD{Here were doing stuff with cubic terms, although we are starting from the unitary gauge action truncated at second order. We should mention this, and say that we just need generic form of cubic terms (I can add a comment about that in sec. 2 where we construct unitary gauge action)}.

In the following we will use these mixing terms to see whether non-Gaussianities in the $\pi$ spectrum can be enhanced by the couplings to $\eta$. We consider the mixing action up to cubic order in the Nambu-Goldstone modes (we will comment on the contribution from the cubic action of the boost breaking sector $S_L$ later):

\begin{equation}\label{equ:cubic_mixing}
\begin{split}
S_{PL} = \int d^4 x \,a^3 &\frac{-\beta_1}{2}\left[ \left(\eta_i - \frac{\partial_i \pi}{a}\right)^2 + 2\dot \pi\frac{\eta^i\partial_i \pi}{a}-2\dot{\pi}\frac{(\partial_i\pi)^2}{a^2}\right] \\
&\frac{-\beta_2}{2}\left[-2\dot \pi \frac{\partial_i \eta^i}{a} +(\partial\pi)^2 \frac{\partial_i \eta^i}{a}  + \dot \pi(3H \eta^2 + 2\eta\dot \eta)\right]\, , \\
\end{split}
\end{equation}
which, in terms of canonical fields and canonical couplings, reads 
\begin{equation}
\label{S_PL_cubic}
\begin{split}
S_{PL} \supset \int d^4 x\, a^3 
	& \beta_1^c H \left[ {\eta_c^i (\partial_i \pi_c/a)} 
	- \frac{\dot \pi_c\eta_c^i(\partial_i \pi_c/a)}{M^2}+\frac{\bar M}{M^4}\dot{\pi_c}\frac{(\partial_i\pi_c)^2}{a^2}\right] \\
&+ a^3\beta_2^c\left[{\dot \pi_c (\partial_i \eta_c^i/a)} -\frac{1}{2}\frac{(\partial\pi_c)^2 (\partial_i \eta_c^i/a)}{M^2}  -\frac{1}{2}\frac{ \dot \pi_c(3H \eta_c^2 + 2\eta_c\dot \eta_c)}{\bar M }\right] \, .
\end{split}
\end{equation}
We emphasize that the terms purely in $\pi$ contained in $S_{PL}$ are possible to write only by breaking boosts as well, not just time-diff. In order words, those operators are impossible to write in the standard EFT of single field inflation. For example, in the standard EFT the term in~$\dot\pi(\partial_i\pi)^2$ is always accompanied by $\dot\pi^2$.

\paragraph{Tree-level Diagrams contributing to $f_{\rm NL}$ :}
We now use the action \eqref{S_PL_cubic} to estimate the $\eta$ induced 3-point function for $\pi$ at horizon crossing, when $\zeta=-H\pi+\ldots$ becomes constant (the $(\d\pi)^3$ vertex in $S_{PL}$ will be discussed at the end of this section). For simplicity, from now on we will restrict to the limit $\beta_i^c \ll 1$, so that we will treat the mixing perturbatively. This implies that the two-point function and the tilt take the standard form as in slow roll inflation
\begin{equation}
\langle\zeta(\vec k_1)\zeta(\vec k_2)\rangle=(2\pi)^3 \delta^{(3)}(\vec k_1+\vec k_2) \frac{H^4}{(-4 \dot H) M_{\rm Pl}^2} \frac{1}{k^3} \ , \qquad n_s-1=4 \frac{\dot H}{H^2}-\frac{\ddot H}{H\dot H}\ ,
\end{equation}
where all quantities are evaluated at horizon crossing. Notice furthermore that the cubic and higher terms in \eqref{S_PL_cubic} are further suppressed by powers of $H^2/M^2$ or $H/\bar M$, and can also be treated perturbatively in the same limit~\footnote{Indeed, this tells us that there is also a regime with $\beta_i^c\gg 1$ where they could still be treated perturbatively. We leave this to future work.}.

From the mixing action \eqref{S_PL_cubic}, we see that both $\beta_1$ and $\beta_2$ contain a mixing term and a $\pi\pi\eta$ interaction, and $\beta_2$ contains an additional $\pi\eta\eta$ vertex. We can therefore consider three different tree-level diagrams that give contributions to $ \left<\pi_c\pi_c\pi_c\right>$. These contributions  translate into a result for $f_{\rm NL}$ by noting that the curvature perturbations $\zeta$ is related to $\pi$ by the relation that we now describe. The curvature perturbation at the time of reheating is related to $\pi$ and $\eta^i$ by the following functional form
\begin{equation}
\zeta = -H\pi +\O(\epsilon \pi^2)+\O((\eta^i-\d_i\pi)^2)\ ,
\end{equation}
The terms in $\epsilon\pi$ come from the relation between $\zeta$ and $\pi$ in single field inflation~\cite{Cheung:2007sv}, while the terms involving $\eta^i$ originate from two types of contributions. The first is associated with the mixing with gravity, while the second is associated with the possible effect of $\eta$ fluctuations at reheating (similarly to the description in the EFT of multifield inflation~\cite{Senatore:2010wk}). Notice that since $\eta^i$ decays outside the horizon, the contribution from these terms at the time of reheating is dominated by modes that are crossing the horizon at the time of reheating itself. Therefore, the contribution is very blue, and negligible at the scale of interest~\footnote{Notice that this contribution enters at loop level, but the diagram is dominated by modes that are crossing the horizon at the time of reheating, the contribution from higher wavenumbers being suppressed by the vacuum prescription~\cite{Senatore:2009cf}.}. We can therefore simply concentrate on a relation between $\zeta$ and $\pi$ which takes the simple form $\zeta=-H\pi$, and use that at horizon crossing $\left<\zeta^2\right> \sim (H^2/M^4)\left<\pi_c \pi_c\right>\sim H^4/M^4 $ (in the last relation we assume that $\pi_c$ is approximately a mass eigenstate, which is true at weak mixing, and use the de Sitter mode function). This implies that at leading order in $\alpha/M_{\rm Pl}^2\ll1$ and $\epsilon\ll1$ 

\begin{equation}
 f_{\rm NL} \equiv \frac{1}{\left<\zeta^2\right>^2} \left<\zeta\zeta\zeta\right> \sim \frac{M^2}{H^5}\left<\pi_c\pi_c\pi_c\right> \, ,
\end{equation}
where, as usual, the time-dependent coefficients are evaluated at horizon crossing. The first diagram one can construct is:
\begin{equation*}
	\includegraphics{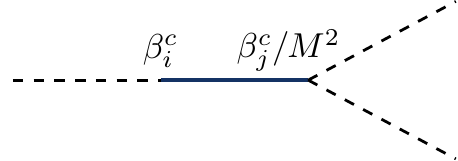}
	\includegraphics{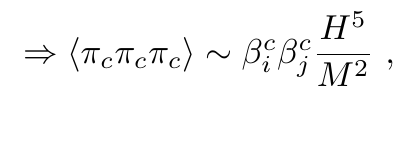}
\end{equation*}
which gives
\begin{equation}
f_{\rm NL}\sim \beta_i^c\beta_j^c \lesssim 1\, .
\end{equation}
Another possibility is to use the $\pi\eta\eta$ vertex in the $\beta_2$ term to construct:
\begin{equation*}
	\includegraphics{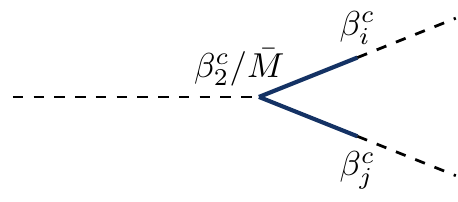}
	\includegraphics{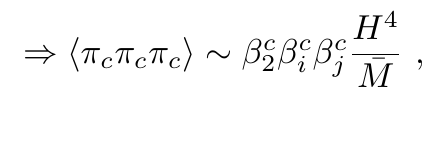}
\end{equation*}

\noindent which gives
\begin{equation}
\label{diagram2}
f_{\rm NL}\sim \beta_i^c\beta_j^c\beta_2^c \frac{M^2}{H\bar M}  \, .
\end{equation}
Finally, one can use the $\eta$ self interaction coming from $S_L$, which will generically be of the form $\alpha^c \eta(\d\eta)^2/\bar M$ (or with derivatives replaced by $H$), with $\alpha^c\sim 1$:
\begin{equation*}
	\includegraphics{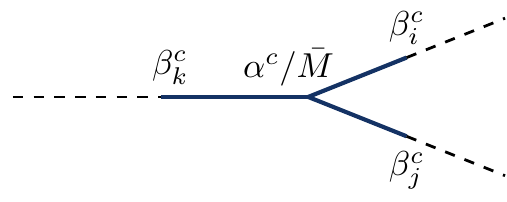}
	\includegraphics{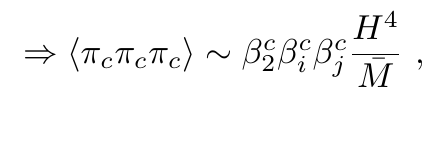}
\end{equation*}

\noindent which gives again
\begin{equation}
f_{\rm NL}\sim \beta_i^c\beta_j^c\beta_k^c \frac{M^2}{H\bar M}  \, .
\label{diagram3}
\end{equation}
The first diagram does not produce large non-Gaussianities, but the others are not necessarily suppressed in the perturbative regime. However, we have not taken into account the stability conditions, which constrain $\beta_1^c$. Forbidding tachyons in the spin 1 sector \eqref{stability_for_12} and requiring positive speed of sound square in the spin 0 sector \eqref{stability_condition_beta} leads to the condition
\begin{align}\label{beta1_conditions}
\frac{-2\bar{M}H}{M^2}\leq\beta_1^c\leq \frac{4M_2^4}{M^2\bar M H}\,.
\end{align}
Since $M_2$ is already a known source of non-Gaussianities \cite{Cheung:2007st}, we will focus here on the case $M_2=0$ where \eqref{beta1_conditions} gives $|\beta_1^c|\lesssim \bar M H/M^2$. This condition implies that the non-Gaussianities from the last two diagrams are bounded by
\begin{align}\label{fNL_estimate}
f_{\rm NL}= (\beta_2^c)^3\frac{M^2}{\bar{M}H}
\lesssim\frac{ (\beta_2^c)^3}{\beta_1^c}\,.
\end{align}
The only way large non-Gaussianities can be produced perturbatively in our setup is if $\beta_1^c$ is strongly suppressed with respect to $\beta_{2}^c$. In the following section we study the naturalness of such a regime.

Note that the action \eqref{S_PL_cubic} also contains a cubic interaction in $\pi$, which could lead to non-Gaussianities of the same shape as single field inflation. Their size is estimated by
\begin{equation}
f_{\rm NL} \sim \frac{\bar M H}{M^2}\beta_1^c = \frac{(\beta_2^c)^3\beta_1^c}{\bar f_{\rm NL}}\, ,
\end{equation}
where in the second step we have used \eqref{fNL_estimate} and called $f_{\rm NL}$ in (\ref{fNL_estimate}) as $\bar f_{\rm NL}$. This contribution is small in the regime of interest $\bar f_{\rm NL}\gtrsim 1$.

%======================================================================%
%============== Naturalness Large mixing regime =======================%
%======================================================================%

\subsection{Naturalness\label{sec:naturalenss}}
In the previous section it was shown that non-Gaussianities \eqref{fNL_estimate} could be large in the perturbative regime only if
\begin{equation}\label{eq:condition}
\beta_1^c \ll (\beta_2^c)^3 \ll 1\, .
\end{equation}
Additionally, we should remember that the $\eta$ fields have a mass equal to
\begin{align}
m_\eta^2=H^2\left(2+\frac{M^2}{\bar{M}H}\beta_1^c\right)\,.
\end{align}
We need to impose this mass to be always less than order $H$, as otherwise the $\eta$ fields will not play any role during inflation, as discussed in sec.~\ref{sec:large_mass}. This implies that
\begin{equation}\label{eq:mass_constr}
\beta_1^c\lesssim\frac{\bar M H}{M^2}\ .
\end{equation}

In this subsection we study if the regime of large $f_{\rm NL}$ of (\ref{eq:condition}), together with the mass constraint from (\ref{eq:mass_constr}), is technically natural. $\beta_1$ controls the correction to the conformal part of the $\eta$ mass; at tree level $\beta_2$ does not generate such a mass, however the cubic vertices
\begin{equation}
\frac{\beta_2^c H}{\bar M} \dot\pi_c \eta_c^2 
\quad \hbox{and} \quad \frac{\beta_1^c H}{M^2}\dot\pi \eta^i\d_i\pi
\end{equation}
generate an $\eta$ mass at one-loop of the form
\begin{equation}\label{rad_mass1}
\delta(m_\eta^2) \sim 
\left(\frac{\beta_2^c H}{\bar M}\right)^2 \int^\Lambda \frac{d^4 k}{(2\pi)^4} \frac{k^2}{k^4}  \sim  
\left(\frac{\beta_2^c H}{\bar M}\right)^2 \Lambda^2 \, ,
\end{equation}
and
\begin{equation}\label{rad_mass2}
\delta(m_\eta^2) \sim 
\left(\frac{\beta_1^c H}{M^2}\right)^2 \int^\Lambda \frac{d^4 k}{(2\pi)^4} \frac{k^4}{k^4}  \sim  
\left(\frac{\beta_1^c H}{M^2}\right)^2 \Lambda^4 \, ,
\end{equation}
respectively, where $\Lambda$ is the cutoff of the EFT. It is most natural to have the cutoff equal to the strong coupling scale, which is the smallest scale suppressing the cubic operators, i.e.\ $\Lambda=\min(M/\sqrt{\beta_2^c},\bar M)$. The regime is then natural if $\delta(m_\eta^2)$ is at most of order of the~$\beta_1$ contribution to the $\eta$ mass $\beta_1/\bar M^2$:  
\begin{equation}
\delta(m_\eta^2) \lesssim \beta_1 /\bar M^2 \quad\Rightarrow \quad\beta_1^c = \frac{\beta_1}{M^2 \bar M H} \gtrsim \frac{\delta(m_\eta^2)\bar M^2}{M^2 \bar M H}\, .
\end{equation}
This is compatible with $\beta_1^c\ll (\beta_2^c)^3$ for both radiatively generated masses as long as
\begin{equation}\label{naturalness}
\frac{H\Lambda^2}{\bar M M^2} \ll \beta_2^c 
\quad \hbox{and} \quad \frac{H\bar M\Lambda^4}{M^6} (\beta_1^c)^2\ll (\beta_2^c)^3 \, .
\end{equation}
We now look at both cases $\Lambda = \bar M$ and $\Lambda = M/\sqrt{\beta_2^c}$ separately.
\paragraph{(1) $\Lambda = \bar M < M/\sqrt{\beta_2^c}$ regime:}
In this case we have
\begin{equation}
\bar M < \frac{M}{\sqrt{\beta_2^c}} \sim \frac{10^{2.5}}{\sqrt{\beta_2^c}} H\, ,
\end{equation}
and the constraints from naturalness \eqref{naturalness} and from the lightness of the $\eta$ fields (\ref{eq:mass_constr}) give, in the regime of interest,
\begin{equation}\label{eq:window1}
10^5 \beta_1^c H\ll \bar M \ll 10^5 \beta_2^c H\, .
\end{equation}
With the condition (\ref{naturalness}) this constraint can always be satisfied by appropriate values for~$\bar M$.
\paragraph{(2) $\Lambda = M/\sqrt{\beta_2^c} < \bar M$ regime:}
In this case we have
\begin{equation}
\bar M > \frac{M}{\sqrt{\beta_2^c}} \sim \frac{10^{2.5}}{\sqrt{\beta_2^c}} H\, ,
\end{equation}
and the constraints from naturalness \eqref{naturalness} and from the lightness of the $\eta$ fields (\ref{eq:mass_constr}) give
\begin{equation}\label{eq:window2}
\frac{1}{(\beta_2^c)^2} H \ll \bar M \ll 10^5 H \frac{(\beta_2^c)^5}{(\beta_1^c)^2}\, .
\end{equation}
Now recall that $f_{\rm NL}\lesssim (\beta_2^c)^3/\beta_1^c$ in (\ref{fNL_estimate}). The smallest window for $\bar M$ is obtained by saturating this bound. Even in this case, we see that there is an appreciable window for $\bar M$ as long as
\begin{equation}
\beta_2^c \gg 10^{-5} f_{\rm NL}^{-2}\, .
\end{equation}
In summary, we find that there exists a parametric window when the conditions in (\ref{eq:condition}) are technically natural.

%======================================================================%
%================= Shape of bispectrum   ==============================%
%======================================================================%

\subsection{Shape of bispectrum}

We now would like to illustrate generic features of the bispectrum in the parameter region $\left|\beta_1^c\right|\lesssim \left|(\beta_2^c)^3\right|\lesssim1$,
under which the mixing interactions can be treated as perturbations and the nonlinearity parameter, $f_{\rm NL}$, could be large enough to be observed.
As given in Eq.~\eqref{eta_parameters},
the mass of the boost Nambu-Goldstone field, $m_\eta$, is modified by the $\beta_1$ coupling as
\begin{align}
m_\eta^2=H^2\left(2+\frac{M^2}{\bar{M}H}\beta_1^c\right)\,.
\end{align}
In all the regime of interests given by (\ref{eq:window1}) and (\ref{eq:window2}) (and with a bare value of $\beta_1^c$ smaller or equal the radiatively generated one~\footnote{It is worth repeating that there is not much room to make the contribution due to $\beta_1$ very large, as otherwise we can integrate the $\eta^i$'s to start with.}), this mass is always equal to $\sqrt{2}H$, apart for the small region of the parameter space where $\bar M\simeq 10^5 \beta_1^c H$ in case  (1) above. We therefore can safely focus on studying the case when $m_\eta\simeq \sqrt{2}H$. This is a very fortunate circumstance for the explicit computations we are going to do next, as this value of the mass of $\eta$ is equal to the conformal mass, a fact that greatly simplifies our analytic expressions.

%
%
%$m_\eta$ is generically of the order Hubble parameter,
%$m_\eta\sim H$.
%In the following discussion,
%for technical simplicity,
%we concentrate on the case
%when $m_\eta$ is the conformal mass, $m_\eta^2=2H^2$.
%Under this assumption,
%we can analytically compute the bispectrum
%without numerical computations.
%Also note that $m_\eta^2=2H^2$ is realized when $\beta_1=0$.
%Although we leave computations of the general mass case for future work,
%qualitative features in the shape of bispectrum we discuss here
%should hold for more general~cases.
%
%In this subsection
%we compute the bispectrum for the second diagram~\eqref{diagram2}.
%For simplicity,
%we assume that $\beta_1=0$.
%Under this assumption, the mass of $\eta$ is the conformal mass,
%so that we can analytically compute the bispectrum.
%Note that,
%on the other hand,
%we need numerical computations for $\beta_1\neq0$.

\paragraph{Hamiltonian in the interaction picture}
As we discussed in Sec.~\ref{subsec:nonG}, the dominant contributions to the bispectrum
are from the two diagrams~\eqref{diagram2} and~\eqref{diagram3}.
Since the latter depends on details of the boost breaking sector, $S_L$,
 for simplicity we focus on the former which is expected to give similar results and is simpler to compute.
The quadratic Hamiltonian for the canonically normalized fields,
$\pi_{c\,{\bf k}}$ and $\sigma_{c\,{\bf k}}$,
can be obtained from the spin 0 action~\eqref{scalar_second}.
Let us choose the free theory part of the Hamiltonian as
\begin{align}
H_{\rm free}^{(2)}&=
\frac{a^2}{2}\int \frac{d^3{\bf k}}{(2\pi)^3}
\Big[
\big(\pi'_{c\,{\bf k}}\pi'_{c\,{\bf -k}}
+c_\pi^2 k^2\,\pi_{c\,{\bf k}}\pi_{c\,{\bf -k}}\big)
+\big(\sigma'_{c\,{\bf k}}\sigma'_{c\,{\bf -k}}
+(c_\sigma^2k^2+a^2m_\eta^2)\sigma_{c\,{\bf k}}\sigma_{c\,{\bf -k}}\big)
\Big]\,,
\end{align}
where we introduced the Hamiltonian as a conjugate to conformal time, $\tau$~(\footnote{Notice that for $M_2=0$, $c_\pi\simeq 1$ in the regime of interest. However, one could in principle make $M_2\neq 0$ and therefore $c_\pi\ll1$, at the cost of introducing an independent source of non-Gaussianities with $f_{\rm NL}\sim 1/c_\pi^2\gg1$. In this case, the signals from the boost sector will become probably subleading, but still potentially detectable.}).
Also, as anticipated, we treat the mixings of $\pi$ and $\sigma$ as interactions.

Canonical quantization then gives
\begin{align}
\pi_{c\,{\bf k}}=u_ka_{\bf k}+u_k^*a_{\bf -k}^\dagger\,,
\quad
\sigma_{c\,{\bf k}}=v_kb_{\bf k}+v_k^*b_{\bf -k}^\dagger\,,
\end{align}
whose Bunch-Davies mode functions, in the de Sitter approximation, are given by
\begin{align}
u_k=\frac{H}{\sqrt{2c_\pi^3k^3}}(1+ic_\pi k\tau)e^{-ic_\pi k\tau}\,,
\quad
v_k=\frac{H}{\sqrt{2c_\sigma^3k^3}}c_\sigma k\tau e^{-ic_\sigma k\tau}\,,
\end{align}
and the commutation relations,
\begin{align}
[a_{\bf k},a^\dagger_{\bf k^\prime}]=(2\pi)^3\delta^{(3)}({\bf k}-{\bf k}^\prime)\,,
\quad
[b_{\bf k},b^\dagger_{\bf k^\prime}]=(2\pi)^3\delta^{(3)}({\bf k}-{\bf k}^\prime)\,.
\end{align}
The interaction Hamiltonian relevant to our computation
is
\begin{align}
H_{\rm mix}^{(2)}&=
a^2\int \frac{d^3{\bf k}}{(2\pi)^3}
\beta_2^c\,k\,\pi_{c\,{\bf k}}'\sigma_{c\,{\bf -k}}
\,,
\\
\nonumber
H_{\rm mix}^{(3)}&=
a^2\prod_i\int \frac{d^3{\bf k}_i}{(2\pi)^3}
(2\pi)^3\delta^{(3)}\Big(\sum_i{\bf k}_i\Big)
\\
&\qquad\qquad
\times
\frac{\beta_2^c}{4\bar{M}}\frac{k_2^2+k_3^2-k_1^2}{k_2k_3}
\pi_{c\,{\bf k}_1}'
\left(
3aH\sigma_{c\,{\bf k}_2}\sigma_{c\,{\bf k}_3}
+\sigma_{c\,{\bf k}_2}'\sigma_{c\,{\bf k}_3}
+\sigma_{c\,{\bf k}_2}\sigma_{c\,{\bf k}_3}'
\right)\,,
\end{align}
where we used the relation,
${\bf k}_2\cdot{\bf k}_3=\frac{1}{2}(k_1^2-k_2^2-k_3^2)$,
for ${\bf k}_1+{\bf k}_2+{\bf k}_3=0$.

\paragraph{Three point functions}
The late time three point correlation functions of $\pi_c$ 
can now be obtained in the in-in formalism from the usual master formula~\cite{Maldacena:2002vr}
\begin{equation}
\left<Q_H(\tau)\right> = 
\left< {0}\, \vphantom{e^{i\int_{-\infty_+}^\tau H_{\rm int}(\tau')d\tau'}} \right|
 \left(\bar T e^{i\int_{-\infty_+}^\tau H_{\rm int}(\tau')d\tau'}\right)Q_I(\tau)\left(T e^{-i\int_{-\infty_-}^\tau H_{\rm int}(\tau')d\tau'}\right)
 \left| \,{0}\vphantom{e^{i\int_{-\infty_+}^\tau H_{\rm int}(\tau')d\tau'}} \right>
\end{equation}

with $Q=\pi_{c\,{\bf k}_1}\pi_{c\,{\bf k}_2}\pi_{c\,{\bf k}_3}$
and $\tau=0$.
For our purposes the exponentials must be expanded up to the third order.
Wick contracting the interaction picture fields,
%
%We now compute three point 
%The three point function of $\pi^c$ is then
%Wick contracting the interaction picture fields,
we have
\begin{align}
\nonumber
&\langle\pi_{c\,{\bf k}_1}\pi_{c\,{\bf k}_2}\pi_{c\,{\bf k}_3}\rangle'
=\frac{H^3}{\sqrt{8c_\pi^9k_1^3k_2^3k_3^3}}2{\rm Re}\bigg[
-i\frac{\beta_2^c}{4\bar{M}}\frac{k_2^2+k_3^2-k_1^2}{k_2k_3}
\int_{-\infty}^0 d\tau\,a^2
\\
&\qquad\quad
\times u_{k_1}'^*
\left(
3aHF_2(\tau)F_3(\tau)
+\widetilde{F}_2(\tau)F_3(\tau)
+F_2(\tau)\widetilde{F}_3(\tau)
\right)
\bigg]
+\text{$2$ permutations}\,,
\end{align}
where
$\langle\pi_{c\,{\bf k}_1}\pi_{c\,{\bf k}_2}\pi_{c\,{\bf k}_3}\rangle=(2\pi)^3\delta^{(3)}\Big(\sum_i{\bf k}_i\Big)\langle\pi_{c\,{\bf k}_1}\pi_{c\,{\bf k}_2}\pi_{c\,{\bf k}_3}\rangle'$.
$F_i(\tau)$ and $\widetilde{F}_i(\tau)$ are defined by
\begin{align}
\nonumber
F_i(\tau)&=-iv_{k_i}^*(\tau)\int_\tau^0 d\tau'\,a^2\beta_2^c\,ku_{k_i}'^*v_{k_i}(\tau')
-iv_{k_i}(\tau)\int_{-\infty}^\tau d\tau'\,a^2\beta_2^c\,ku_{k_i}'^*v_{k_i}^*(\tau')
\\
%\nonumber
&\quad
+iv_{k_i}^*(\tau)\int_{-\infty}^\infty d\tau'\,a^2\beta_2^c\,ku'_{k_i}v_{k_i}(\tau')\,,
%\\
%&=\frac{\beta_2^c}{2}\sqrt{\frac{c_\pi}{c_\sigma}}\left[
%v_{k_i}^*(\tau)\left(\frac{e^{i(c_\pi-c_\sigma)k_i\tau}-1}{c_\pi-c_\sigma}
%-\frac{1}{c_\pi+c_\sigma}\right)
%-v_{k_i}(t)\frac{e^{i(c_\pi+c_\sigma)k_i\tau}}{c_\pi+c_\sigma}
%\right]\,,
\\
\nonumber
\widetilde{F}_i(\tau)&=-iv_{k_i}'^*(\tau)\int_\tau^\infty d\tau'
\,a^2\beta_2^c\,ku_{k_i}'^*v_{k_i}(\tau')
-i\dot{v}_{k_i}(\tau)\int_{-\infty}^\tau d\tau'\,a^2\beta_2^c\,ku_{k_i}'^*v_{k_i}^*(\tau')
\\
%\nonumber
&\quad
+iv_{k_i}'^*(\tau)\int_{-\infty}^\infty d\tau'\,a^2\beta_2^c\,ku'_{k_i}v_{k_i}(\tau')\,.
%\\
%&=\frac{\beta_2^c}{2}\sqrt{\frac{c_\pi}{c_\sigma}}\left[
%v_{k_i}'^*(\tau)\left(\frac{e^{i(c_\pi-c_\sigma)k_i\tau}-1}{c_\pi-c_\sigma}
%-\frac{1}{c_\pi+c_\sigma}\right)
%-v'_{k_i}(\tau)\frac{e^{i(c_\pi+c_\sigma)k_i\tau}}{c_\pi+c_\sigma}
%\right]\,.
\end{align}
More explicitly,
\begin{align}
F_i(\tau)&=\frac{\beta_2^c}{2}\sqrt{\frac{c_\pi}{c_\sigma}}\left[
v_{k_i}^*(\tau)\left(\frac{e^{i(c_\pi-c_\sigma)k_i\tau}-1}{c_\pi-c_\sigma}
-\frac{1}{c_\pi+c_\sigma}\right)
-v_{k_i}(\tau)\frac{e^{i(c_\pi+c_\sigma)k_i\tau}}{c_\pi+c_\sigma}
\right]
\,,
\\
\widetilde{F}_i(\tau)&=\frac{\beta_2^c}{2}\sqrt{\frac{c_\pi}{c_\sigma}}\left[
v_{k_i}'^*(\tau)\left(\frac{e^{i(c_\pi-c_\sigma)k_i\tau}-1}{c_\pi-c_\sigma}
-\frac{1}{c_\pi+c_\sigma}\right)
-v'_{k_i}(\tau)\frac{e^{i(c_\pi+c_\sigma)k_i\tau}}{c_\pi+c_\sigma}
\right]\,.
\end{align}

The three point function is now reduced to the form,
\begin{align}
\nonumber
\langle\pi_{c\,{\bf k}_1}\pi_{c\,{\bf k}_2}\pi_{c\,{\bf k}_3}\rangle'
&=\frac{H^3}{k_1^2k_2^2k_3^2}\frac{K^3}{k_1k_2k_3}
\frac{1}{64c_\pi^4c_\sigma^3}
(\beta_2^c)^3\frac{H}{\bar{M}}
\kappa_1^2\left(\kappa_2^2+\kappa_3^2-\kappa_1^2\right)
\\
\nonumber
&
\quad\times{\rm Im}\left[
\int_0^\infty dx\,
e^{-ic_\pi \kappa_1x}
\left(
3f_2(x)f_3(x)
+\widetilde{f}_2(x)f_3(x)
+f_2(t)\widetilde{f}_3(x)
\right)
\right]
\\*
\label{pipipi}
&\quad
+\text{$2$ permutations}\,,
\end{align}
where we introduced $x=-(k_1+k_2+k_3)\tau$ and $\kappa_i=\frac{k_i}{k_1+k_2+k_3}$.
The functions $f_i$ and $\widetilde{f}_i$ are of the form
\begin{align}
f_i(x)&=\sqrt{c_\pi c_\sigma}
\left[
\frac{e^{-ic_\pi \kappa_ix}+ e^{-ic_\sigma\kappa_ix}}{c_\pi+c_\sigma}
-\frac{e^{-ic_\pi\kappa_i x}- e^{-ic_\sigma\kappa_ix}}{c_\pi-c_\sigma}
\right]\,,
\\
\nonumber
\widetilde{f}_i(x)&=\sqrt{c_\pi c_\sigma}
\left[
-\frac{1}{c_\pi+c_\sigma}\Big(
(1+ic_\sigma\kappa_ix)e^{-ic_\pi \kappa_ix}+
(1-ic_\sigma\kappa_ix)e^{-ic_\sigma\kappa_ix}\Big)\right.
\\
&\qquad\qquad\,\,\,
\left.+\frac{1-ic_\sigma\kappa_ix}{c_\pi-c_\sigma}\left(e^{-ic_\pi\kappa_i x}- e^{-ic_\sigma\kappa_ix}\right)
\right]\,.
\end{align}
Note that for $c_\pi=c_\sigma$ we have
\begin{align}
f_i(x)=
(1+ic_\pi\kappa_ix)e^{-ic_\pi \kappa_ix}
\,,
\quad
\widetilde{f}_i(x)=
-\left(1+ic_\pi\kappa_ix+c_\pi^2\kappa_i^2x^2\right)
e^{-ic_\pi \kappa_ix}\,.
\end{align}
%The integration in~\eqref{pipipi}
%is performed in the Mathematica note.

\paragraph{Shape function}
We then evaluate the shape function,
\begin{align}
S(k_1,k_2,k_3)&=\frac{k_1^2k_2^2k_3^2}{(2\pi)^4\mathcal{P}_\zeta^2}
\langle\zeta_{{\bf k}_1}\zeta_{{\bf k}_2}\zeta_{{\bf k}_3}\rangle'
\quad
{\rm with}
\quad
\mathcal{P}_\zeta=\frac{k^3}{2\pi^2}\langle\zeta_{{\bf k}}\zeta_{-{\bf k}}\rangle'\,.
\end{align}
Using the linear order relation $\zeta=-H\pi$, we have
\begin{align}
\langle\zeta_{{\bf k}_1}\zeta_{{\bf k}_2}\zeta_{{\bf k}_3}\rangle'
=-\frac{H^3}{M^6}\langle\pi^c_{{\bf k}_1}\pi^c_{{\bf k}_2}\pi^c_{{\bf k}_3}\rangle'\,,
\end{align}
and then
\begin{align}
\nonumber
S(k_1,k_2,k_3)
%&=\frac{k_1^2k_2^2k_3^2}{4}
%\langle\zeta_{{\bf k}_1}\zeta_{{\bf k}_2}\zeta_{{\bf k}_3}\rangle'\left(\frac{k^3}{\langle\zeta_{{\bf k}}\zeta_{-{\bf k}}\rangle'}\right)^2
%\\
%&=\frac{k_1^2k_2^2k_3^2}{4}
%\langle\zeta_{{\bf k}_1}\zeta_{{\bf k}_2}\zeta_{{\bf k}_3}\rangle'\left(\frac{2M^4c_\pi^3}{H^4}\right)^2
%\\
%&=-\frac{M^2}{H^5}c_\pi^6k_1^2k_2^2k_3^2
%\langle\pi^c_{{\bf k}_1}\pi^c_{{\bf k}_2}\pi^c_{{\bf k}_3}\rangle'
%\\
&=-\frac{M^2}{\bar{M}H}(\beta_2^c)^3\frac{c_\pi^2}{64c_\sigma^3}\frac{\kappa_1^2\left(\kappa_2^2+\kappa_3^2-\kappa_1^2\right)}{\kappa_1\kappa_2\kappa_3}
\\
\nonumber
&
\quad\times{\rm Im}\left[
\int_0^\infty dx\,
e^{-ic_\pi \kappa_1x}
\left(
3f_2(x)f_3(x)
+\widetilde{f}_2(x)f_3(x)
+f_2(x)\widetilde{f}_3(x)
\right)
\right]
\\*
\label{shape_1}
&\quad
+\text{$2$ permutations}
\end{align}
at the leading order in $\beta_2^c$.
The remaining integral can be computed analytically, and its expression is given in Appendix~\ref{App:shape}.

\begin{figure}[t]
\begin{center}
\includegraphics[width=150mm, bb=0 0 816 218]{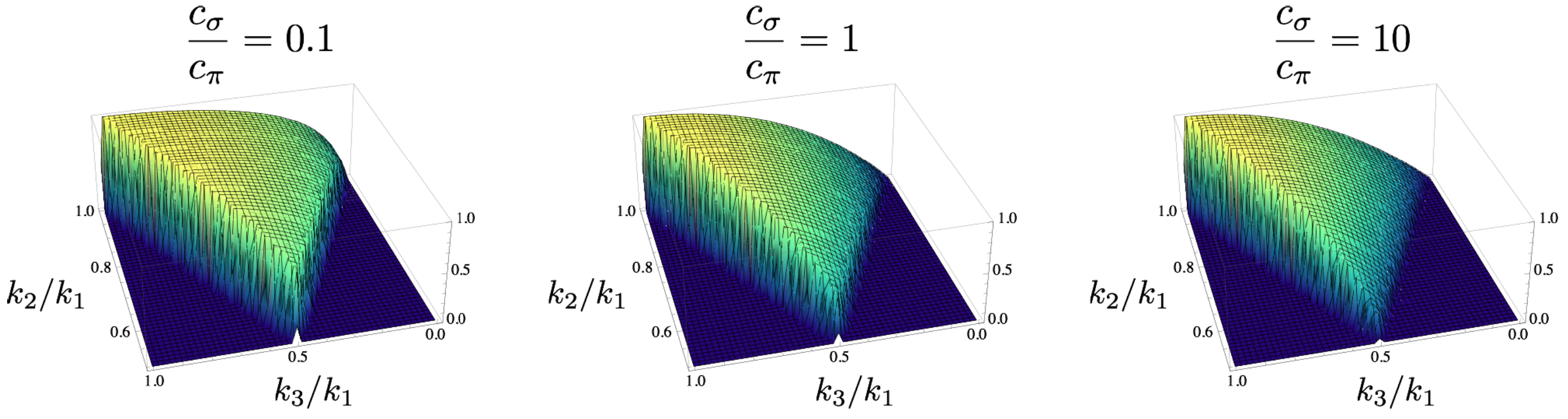}
\end{center}
\vspace{-4mm}
 \caption{The shape function $S(k_1,k_2,k_3)$
 for $r=c_\sigma/c_\pi=0.1,1,10$.
 The plot is normalized such that $S(k,k,k)=1$.
 We notice that the peak is at the equilateral configuration
 and the slope becomes flatter as $r$ gets smaller.}
\label{shape}
\end{figure}

\begin{figure}[t]
\begin{center}
\includegraphics[width=80mm, bb=0 0 292 196]{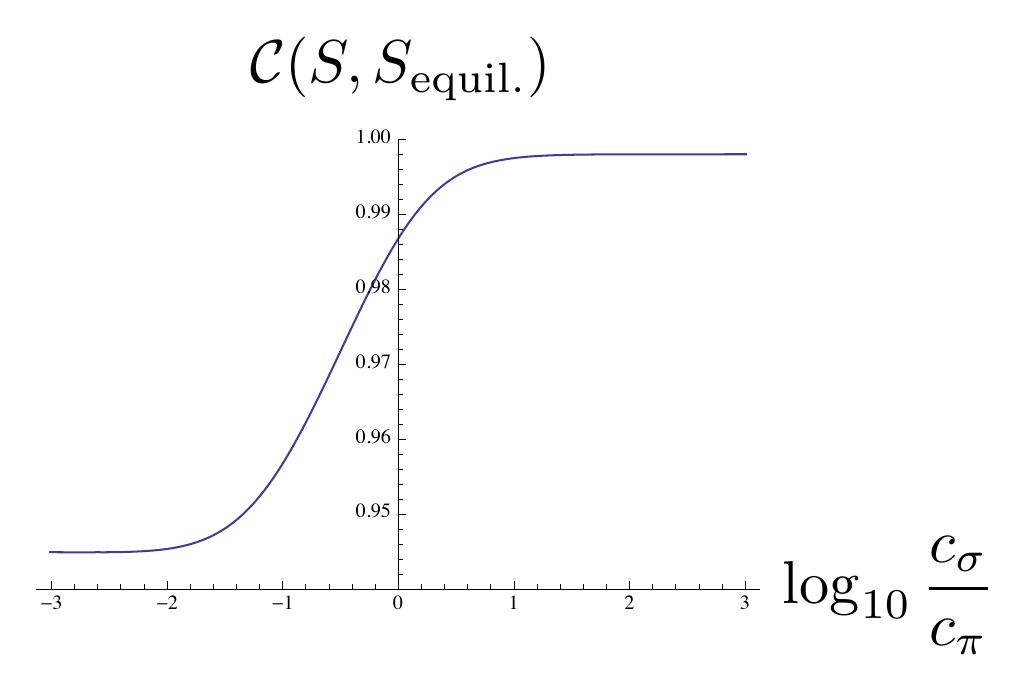}
\end{center}
\vspace{-6mm}
 \caption{The cosine parameter, $\mathcal{C}(S,S_{\rm equil.})$, vs $r=c_\sigma/c_\pi$.
}
\label{cosine}
\end{figure}

\paragraph{Cosine parameter}
As depicted in Fig.~\ref{shape},
the shape function has a peak at the equilateral configuration~\footnote{For observational purposes, we expect that $c_\pi\simeq 1$, and so $c_\sigma\lesssim c_\pi$. However, in the figures, for explanatory purposes, we also plot the regime $c_\sigma\gtrsim c_\pi$.}.
Also the slope becomes flatter
as the ratio, $r=c_\sigma/c_\pi$, of sound speeds gets smaller.
To characterize how the shape function is similar to the equilateral form,
let us compute the cosine parameter~\cite{Babich:2004gb}
for the shape function~\eqref{shape_1}
and the equilateral shape~\cite{Creminelli:2005hu} function,
\begin{align}
S_{\rm equil.}(k_1,k_2,k_3)=
\left(\frac{k_1}{k_2}+\text{5 permutations}\right)-\left(\frac{k_1^3}{k_1k_2k_3}+\text{2 permutations}\right)-2\,.
\end{align}
For two shape functions, $S_1(k_1,k_2,k_3)$ and $S_2(k_1,k_2,k_3)$,
we introduce the inner product~as
\begin{align}
\label{equilateral_shape}
S_1\cdot S_2 =\int_0^1 dx \int_{1-x}^{1}dy \,S_1(1,x,y)S_2(1,x,y)\,.
\end{align}
The cosine parameter, $\mathcal{C}(S_1,S_2)$,
is then defined by
\begin{align}
\mathcal{C}(S_1,S_2)=\frac{S_1\cdot S_2}{(S_1\cdot S_1)^{1/2}(S_2\cdot S_2)^{1/2}}
\,,
\end{align}
where note that $\mathcal{C}(S,S)=1$.
In Fig.~\ref{cosine},
we plot the cosine parameter, $\mathcal{C}(S,S_{\rm equil.})$,
for our shape function~\eqref{shape_1} and the equilateral template~\eqref{equilateral_shape}.
As was suggested by Fig.~\ref{shape},
the cosine parameter decreases as the ratio, $r=c_\sigma/c_\pi$, of sound speeds gets smaller.
Note that the cosine parameter approaches to $0.945$ and $0.987$ in the limits $c_\pi\gg c_\sigma$ and $c_\pi\lesssim c_\sigma$, respectively.
Asymptotic behaviors of the full bispectrum are also given in Appendix~\ref{App:shape}.

\paragraph{Nonlinearity parameter $f_{\rm NL}$}
The $f_{\rm NL}$ parameter can be computed as
\begin{align}
f_{\rm NL}=\frac{10}{9}S(k,k,k)=\frac{5}{216}\frac{36+198r+281r^2+140r^3+20r^4}{c_\pi^2r^2(1+r)^2(2+r)^2(1+2r)^2}\frac{M^2}{\bar{M}H}(\beta_2^c)^3
\quad
{\rm with}
\quad
r=\frac{c_\sigma}{c_\pi}\,.
\end{align}

If we use $M^4=2\epsilon M_{\rm Pl}^2H^2$, which is valid for $M_2=0$ in the decoupling regime,
we can rewrite it as
\begin{align}
f_{\rm NL}=\frac{5\sqrt{2}}{216}\frac{36+198r+281r^2+140r^3+20r^4}{c_\pi^2r^2(1+r)^2(2+r)^2(1+2r)^2}\frac{\epsilon^{1/2}M_{\rm Pl}}{\bar{M}}(\beta_2^c)^3\,,
\end{align}
which scales as anticipated in (\ref{fNL_estimate}). In particular, in the parametric regime highlighted in Sec.~\ref{sec:naturalenss}, $f_{\rm NL}$ can be parametrically larger than one, up to order $(\beta_2^c)^3/\beta_1^c$. The numerical prefactor is $\simeq 0.07$ for $r\to 1$, or $\simeq 64/r^2$ for $r\ll1$, which highlights the common enhancement of non-Gaussianities for small speed of sounds.

\paragraph{Squeezed limit}

Finally, the squeezed limit of the shape function is given by
\begin{align}
\lim_{k_1\ll k_2} S(k_1,k_2,k_2)=\frac{4c_\pi^5+20c_\pi^4 c_s+25c_\pi^3c_s^2+7c_\pi^2c_s^3}{32c_\sigma^3(c_\pi+c_\sigma)^4}\frac{M^2}{\bar{M}H}(\beta_2^c)^3\frac{k_1}{k_2}\,,
\end{align}
which can also be expressed in terms of the three point function as
\begin{align}\label{eq:squeezed}
\lim_{k_1\ll k_2}\langle\zeta_{\vec k_1}\zeta_{\vec k_2}\zeta_{\vec k_3}\rangle' \propto \langle\zeta_{\vec k_1}^2\rangle'  \langle\zeta_{\vec k_2}^2\rangle' \left(\frac{k_1}{k_2}\right)^{2 }\,.
\end{align}
Here it should be emphasized that this scaling behavior is different from the original quasi-single field model~\cite{Chen:2009zp} with the conformal mass $m=\sqrt{2}H$, where we have
\begin{align}
\lim_{k_1\ll k_2}\langle\zeta_{\vec k_1}\zeta_{\vec k_2}\zeta_{\vec k_3}\rangle' \propto \langle\zeta_{\vec k_1}^2\rangle'  \langle\zeta_{\vec k_2}^2\rangle' \left(\frac{k_1}{k_2}\right)\,.
\end{align}
As studied in~\cite{Noumi:2012vr}, the scaling behavior depends on the form of cubic interactions in general. We can also reproduce the same scaling behavior by following the same argument there.

Although the signal is very small in our squeezed limit in (\ref{eq:squeezed}), comparable to the one obtained in models with only one degree of freedom, such as in the case of the so-called equilateral~\cite{Creminelli:2005hu} and orthogonal~\cite{Senatore:2009gt} shapes, it nevertheless could show up in large scale structures surveys as a bias term with a peculiar functional form (see for example~\cite{Senatore:2014eva}).

%
%

%--------------------------------------------------------------------------------------
% SECTION: ACKNOWLEDGMENTS
%--------------------------------------------------------------------------------------
\section*{Acknowledgments}

We would like to thank Riccardo Penco for discussions. T.N. thanks Stanford Institute for Theoretical Physics at Stanford University, where a part of this work was done. T.N. is partially supported by a grant from Research Grants Council of the Hong Kong Special Administrative Region, China (HKUST4/CRF/13G), Special Postdoctoral Researcher Program at RIKEN and the RIKEN iTHES project. L.S. is partially supported by DOE Early Career Award DE-FG02-12ER41854.

%%%%%%%%%%%%%%%%%%%%%%%%%%%%%%%%%%%%%%%%%%%%%%%%%%%%%%%%%%%%%%%%%%%%%%%%
%======================================================================%
%=========================== APPENDIX =================================%
%======================================================================%
%%%%%%%%%%%%%%%%%%%%%%%%%%%%%%%%%%%%%%%%%%%%%%%%%%%%%%%%%%%%%%%%%%%%%%%%

\appendix
\section*{Appendices}

%======================================================================%
%====================== Pure Boost ====================================%
%======================================================================%

\section{Pure boost}
It is interesting to note that even if time diffs are unbroken, the boost sector alone $S=S_{EH} + S_L$ can source FRW. Indeed the background equations of motion \eqref{background} have a solution for $c=0$ given by~\footnote{This is related to the observation that for an FRW background, the stress tensor of the boost-breaking sector is proportional to the Einstein tensor $T^\eta_{\mu\nu} \propto G_{\mu\nu}$ \cite{Jacobson:2008aj}. In equation \eqref{pure_boost_sol} we are tuning the $\alpha$'s such that they are equal.}
\begin{equation}\label{pure_boost_sol}
3\alpha_2 + \alpha_3 = -2 M_{\rm Pl}^2 \, , \quad \Lambda = 0 \, .
\end{equation}
The fluctuations around FRW are now somewhat exotic because of the absence of the Nambu-Goldstone mode $\pi$ for time diffs. The fact that $\alpha\sim M_{\rm Pl}^2$ also has interesting consequences: it suggests that the $\eta$ strong coupling scale is of the order $M_{\rm Pl}$. In addition, mixing with gravity, which is of order $\alpha/M_{\rm Pl}^2$, is not suppressed, so there is no decoupling limit and the constrained variables should be carefully integrated out. Notice also that this spacetime solution is highly tuned: as soon as we violate (\ref{pure_boost_sol}), the background solution becomes Minkowski spacetime (or de Sitter by making $\Lambda\neq 0$).

In section \ref{app:coset} we show how the boost sector can be equivalently obtained from a coset construction. In section \ref{pureboostdof} we solve the constraint equations, and show that the scalar component $\eta_i^\parallel = \d_i \sigma$ is generically frozen. Surprisingly, the propagating degrees of freedom are therefore the graviton and a spin 1 mode $\eta_i^\bot$. 

%====================== CC for pure boost =============================%

\subsection{Coset Construction for Boost Sector}\label{app:coset}

In this section we construct the boost sector $S=S_{EH} + S_L$ using the coset construction~\cite{Coleman:1969sm,Callan:1969sn} for non-linear realizations of spacetime symmetries \cite{Volkov:1973vd,Ogievetsky:1974} in curved space \cite{Delacretaz:2014oxa}. A nice review of coset techniques can be found in \cite{Goon:2012dy}. 
This is the curved space extension of ``framids'', which were recently discussed in a condensed matter setting in \cite{Nicolis:2015sra}. The $\eta$ field arises as the Nambu-Goldstone modes associated with broken boosts in the symmetry breaking pattern
\begin{equation}
P_a ,\, J_i ,\, K_i \to P_a ,\, J_i \, .
\end{equation}
The coset space can be parametrized with the representative $\Omega = e^{x^a P_a}e^{\eta^i K_i}$. The Maurer-Cartan form in curved space is then given by
\begin{equation}\label{MC_form}
\begin{split}
\Omega^{-1} (d + \tilde e^a P_a + \frac{1}{2} \omega^{ab} J_{ab})\Omega 
	& \,= e^{-\eta K} (d + e^a P_a + \frac{1}{2} \omega^{ab} J_{ab})e^{\eta K} \\
	&\, = e^a \Lambda_a{}^b P_b + \frac{1}{2}[\Lambda^T (d + \omega) \Lambda]^{ab} J_{ab}
\end{split}
\end{equation}
where $\Lambda^a{}_b (\eta) = (e^{\eta^iK_i})^a{}_b$ is a boost in the 4-vector representation, and $\tilde e^a = e^a - dx^a - \omega^{ab}x_b$ ($e^a$ is the physical vierbein, related to the metric by $g_{\mu\nu}=e_\mu^ae_\nu^b\eta_{ab}$). In this formalism, invariant Lagrangians are built from the coefficients of the generators in the Maurer-Cartan form. The coefficient of $P_\mu$ (the ``coset vielbein'') and that of $J_{0i}$ (coset covariant derivative of $\eta$) can be read off from \eqref{MC_form}
\begin{equation}
E^b = dx^\mu E_\mu^b = dx^\mu e^a_\mu \Lambda_a{}^b  \, , \quad
\mathcal D_a \eta^i = E_a^\mu \Lambda_b{}^i (\eta^{bc}\d_\mu + \omega_\mu^{bc})\Lambda^0{}_c\, .
\end{equation}
The most general action is then given by
\begin{equation}
S = \int d^4 x |\det E|  \, \mathcal L (\mathcal D_a \eta^i, \ldots )
\end{equation}
where $\mathcal L$ is a function of the coset building blocks that is invariant under the unbroken group --  here rotations -- and the dots denote higher derivative terms. Notice that $|\det E| = |\det e| = \sqrt{-g}$. Up to quadratic order the terms allowed in $\mathcal L$ are
\begin{equation}
\mathcal D_i \eta^i \, , \quad
(\mathcal D_i \eta_j - \mathcal D_j \eta_i)^2 \, , \quad
(\mathcal D_i \eta^i)^2 \, , \quad
(\mathcal D_a \eta^j)^2 \, , \quad
(\mathcal D_0 \eta^i)^2 \, .
\end{equation}
The first one gives a total derivative: 
\begin{equation}
\begin{split}
\mathcal D_i \eta^i 
	& =  e_a^\mu \Lambda^a{}_i \Lambda_b{}^i (\eta^{bc}\d_\mu + \omega_\mu^{bc})\Lambda^0{}_c \\ 
	& = e_a^\mu \d_\mu \Lambda^{0a} + e_a^\mu \omega_\mu^{ab}\Lambda_b{}^0 \\
	& = e_a^\mu \d_\mu \Lambda^{0a} + \Lambda^{a0} \nabla_\mu e_a^\mu = \nabla_\mu Q^\mu \, , 
\end{split}
\end{equation}
where in the last line we expressed the torsion free spin-connection in terms of the vielbein $\omega_{\mu, \, ab} = e_a{}^\nu \nabla_\mu e_{b\nu}$, and $Q_\mu$ is defined as in the main text \eqref{Q_def} as $Q_\mu = \Lambda^0{}_a e_\mu^a$. Similar calculations show that the others give rise to the four $\alpha_i$ terms in \eqref{S_stu}:
\begin{subequations}
\begin{align}
(\mathcal D_{[i} \eta_{j]} )^2
	&= (\nabla_{[\mu}Q_{\nu]})^2 + \frac{1}{2}(Q^\mu \nabla_\mu Q_\nu)^2\, ,\\
(\mathcal D_{i} \eta^{i} )^2
	&= ( \nabla_\mu Q^\mu)^2 \, ,\\
(\mathcal D_a \eta^j)^2 
	&= (\nabla_\mu Q_\nu)^2 \, , \\
(\mathcal D_0 \eta^i )^2 
	&= (Q^\mu \nabla_\mu Q_\nu)^2\, .
\end{align}
\end{subequations}
The coset construction exactly reproduces the gauge invariant action with St\"uckelberg fields \eqref{S_stu}:
\begin{equation}
S_L = \int d^4 x  |\det E| \left[
	- {\alpha_1}(\mathcal D_{[i} \eta_{j]})^2
	- \frac{\alpha_2}{2}(\mathcal D_i \eta^i)^2
	- \frac{\alpha_3}{2}(\mathcal D_a \eta^j)^2
	- \frac{\alpha_4-\alpha_1}{2}(\mathcal D_0 \eta^i)^2
\right] \, .
\end{equation}
%

%================ DoFs  ====================%

\subsection{Degrees of Freedom in pure boost}\label{pureboostdof}

The action $S=S_{EH}+S_L$ contains constrained ADM variables $\delta N$ and $N_i$ (see appendix \ref{app:ADM} for details). In this section we solve the constraint equations arising from the second order action \eqref{S_ADM}, with ${\cal L}_P={\cal L}_{PL}=0$, to establish an action containing only the propagating degrees of freedom $\eta$ and $\gamma$. The solution to the constraint equations are
\begin{subequations}
\begin{align}
\delta N 
	&= a{\dot \sigma} + a{H\sigma}\left[1- \frac{3\alpha_2+\alpha_3}{\alpha_1-\alpha_4+\alpha_3}\right]\, ,\\
N_i  
	&= -a\d_i\sigma + \frac{\alpha_3}{M_{\rm Pl}^2 - \alpha_3} a \eta_i^\bot  \, .
\end{align}
\end{subequations}
where $\eta_i =\eta_i^\bot +  \d_i \sigma$. Returning to the action, this gives
\begin{equation*}\label{S2_pureboost}
\begin{split}
&S^{(2)}
	=\int d^4 x \, a^3 \left[
\frac{M_{\rm \rm Pl}^2 - \alpha_3}{8}\dot\gamma_{ij{\bf}}\dot\gamma_{ij{\bf}}
-\frac{M_{\rm \rm Pl}^2}{8}\,(\d_k\gamma_{ij})^2/a^2\right] \\
	&  +\int d^4 x \, a^3 \left[ \frac{\alpha_1+\alpha_3-\alpha_4}{2}(\dot\eta_{\bot i}^2 -2 H\eta_{\bot i}^2)  - \frac{1}{2}\left(\alpha_1 - \frac{\alpha_3^2}{3 (\alpha_2 +\alpha_3)}\right) (\d_i\eta_{\bot j})^2/a^2 + \frac{3}{2}\alpha_2\dot H (\eta_{\bot i})^2\right]\\
	& + \int d^4 x \, a^3 M_{\rm Pl}^2 H^2 \mu(\alpha_i/M_{\rm Pl}^2)\; (\eta_i^\parallel)^2 \, ,
\end{split}
\end{equation*}
where $\mu$ is a function of $\alpha_i/M_{\rm Pl}^2$ with order one coefficients, so that $\mu$ itself is order one.
As usual the $\alpha$'s will have to satisfy inequalities for stability. The most surprising feature of \eqref{S2_pureboost} is that $\eta_i^\parallel$ does not have a kinetic term, but it does have a mass term in general. As a consequence, the spin 0 mode is frozen, and the only propagating degrees of freedom are the graviton and the spin 1 mode $\eta_i^\bot$. Interestingly, there are some special values of $\alpha_i$ such that $\mu=0$ and therefore the $\eta_i^\parallel$ mass term vanishes. In such a case the cubic action is crucial in determining the dynamics; we leave this issue for future work.

Although this setup has stable fluctuations around FRW, it will not give an inflationary model compatible with current observations. Indeed, the only scalar operators in this model are composite operators such as $\eta_i^2$; therefore in order to have a reasonable inflationary model with scale invariant and quasi Gaussian spectrum, one needs to add a scalar field by hand to the picture, making this a multifield model. We leave the study of this model to future work.

%======================================================================%
%================ Goldstone-Gravity mixing and ADM ====================%
%======================================================================%

\section{Unmixing Goldstones and Gravity}\label{app:ADM}

The goal of this section is to show that the action for boost breaking inflation \eqref{S_stu}, which contains mixing between the Nambu-Goldstone modes and the metric, reduces to the decoupling limit action \eqref{S_DL} at energies $E\gg \epsilon^{1/2}H$ as long as the conditions \eqref{DL2} are satisfied. Since the Nambu-Goldstone modes are spin 0 and 1, they only mix with the constrained components of the metric, so the action can be unmixed by solving the constraint equations.
It is most convenient to study constraints using ADM variables, which we introduce at the level of the vierbein (see e.g. \cite{Hinterbichler:2012cn})
\begin{equation*}
e_\mu^a =
\left( \begin{array}{cc}
N & N^i e_i^m  \\
0 & e_i^m \end{array} \right) 
\hspace{7pt}
\Rightarrow
\hspace{7pt}
g_{\mu\nu} = 
\left( \begin{array}{cc}
-N^2 + N^i N_i & N_i  \\
N_i & h_{ij} \end{array} \right) 
\hspace{7pt}
\hbox{or}
\hspace{7pt}
g^{\mu\nu} = 
\left( \begin{array}{cc}
-1/N^2  & N^i/N^2  \\
N^i/N^2 & h^{ij} - N^iN^j/N^2 \end{array} \right) 
\end{equation*}
where $h_{ij}=\delta_{mn}e_i^m e_j^n$ and $N_i=h_{ij}N^j$. Here $i,j...=1,2,3$ denote spatial diff indices and $m,n,...=1,2,3$ denote spatial local Lorentz indices. Local Lorentz invariance was partially fixed in order to obtain $e_\mu^0=\delta_\mu^0 N$ (this is allowed as $e_i^0$ is not a constrained variable). It is also useful to define the extrinsic curvature of equal time slices: $K_{ij} = \frac{1}{2N}(\dot h_{ij} - \nabla_i N_j - \nabla_j N_i)$. In terms of these variables we have for example
\begin{align*}
\sqrt{-g}
	&= N\sqrt{h} \ , \\
R
	&= R^{(3)} - K^2 + K^i{}_j K^j{}_i+{\rm \; total\;derivative}\ , \\
g^{\mu\nu}\d_\mu(t+\pi)\d_\nu(t+\pi)
	&=  -\frac{1}{N^2} (1+ \dot \pi - N^i\d_i \pi)^2 + h^{ij}\d_i\pi\d_j\pi\ .
%(\nabla_\mu (\Lambda^{\bar 0}{}_a e_\nu^{a}))^2 
	%&\sim   2 H N^i\d_i N - 2 H^2 N_i N^i - (\d_i N/a)^2 +  \dot \eta_i^2  -(\d_i\eta_j/a )^2 \\
	%& \quad + H h^{ij} \d_i \eta_j + H^2 N^i \eta_i + H N \d_i \eta_i  + \O (\delta^3)\nonumber
\end{align*}
The other terms that appear in the action \eqref{S_stu} can be worked out similarly. The full quadratic action in terms of ADM variables and Nambu-Goldstone fields is 
\begin{subequations}\label{S_ADM}
\begin{align}
S &= 
	\int d^4 x \sqrt{h} (\mathcal L_{EH} + \mathcal L_P + \mathcal L_L + \mathcal L_{PL})\\
\mathcal L_{EH}  &= \label{equ:SEH_ADM}
	\frac{1}{2}M_{\rm Pl}^2( NR^{(3)} -  N K^2 + N K^i{}_j K^j{}_i) \\
\mathcal L_P  &= \label{equ:SP_ADM}
	  - N\Lambda  +\frac{c}{N} (1+ \dot \pi - N^i\d_i \pi)^2   - c \, (\d_i\pi/a)^2 + 2 M_2^4 (\delta N - \dot \pi)^2 \vphantom{\frac{1}{2}}\\
\mathcal L_L  &=  \nonumber
	\frac{\alpha_1}{2}\left[(\dot \eta_i + H \eta_i - \d_i N/a)^2 - \frac{1}{2}(\d_i\eta_j - \d_j\eta_i)^2/a^2\right]
		-\frac{\alpha_2}{2}\left[(\d_i\eta_i/a + 3\dot H \pi - \delta K)^2\right] \\
		&\nonumber \quad
		 +\frac{\alpha_3}{2}\left[ - NK_i{}^jK_j{}^i + (\d_iN/a- \dot \eta_i)^2  -2 H^2\eta^2 -(\d_i\eta_j/a)^2\vphantom{\gamma_{ij}^k}+ 2a\, \delta (N K^{ij})\d_i\eta_j \vphantom{\gamma_{ij}^k}   \right] \\
		&\quad\label{equ:SL_ADM} 
		-\frac{\dot \alpha_3}{2}H \frac{\eta_i\d_i\pi}{a}-\frac{\alpha_4}{2}\left[\dot \eta_i + H\eta_i - \d_i N /a\right]^2\\
%		&\hspace{65pt} \nonumber
%		\left.-\frac{\alpha_4}{2}\left[\dot \eta_i + H\eta_i - (\d_i N + HN_i)/a\right]^2
%		\vphantom{\frac{1}{1}}\right\} \\
\mathcal L_{PL}  &= \label{equ:SPL_ADM}
	\frac{-\beta_1}{2}\left(\eta_i - \frac{\d_i\pi}{a}\right)^2
	+ \beta_2 (\dot \pi - \delta N ) \left(\frac{\d_i\eta_i}{a} - \delta K + 3\dot H \pi\right)
\end{align}
\end{subequations}
where all time-dependent coefficients ($c,\, \Lambda,\, M_2,\, \alpha_i,\, \beta_i$) are evaluated at $t+\pi$, and a few terms that are subleading in the limit \eqref{DL2} and in the slow roll expansion were dropped.

We now have to solve the constraint equations $\delta S/\delta N=0$, $\delta S/\delta N_i=0$ in terms of the Nambu-Goldstone fields, and plug the solutions back into the action. The spin 1 part $N_i^\bot$ can be found fairly easily by solving the transverse equation $(\delta S/\delta N_i)_\bot$. The solution is
\begin{equation}
N_i^\bot = \frac{\alpha_3}{M_{\rm Pl}^2 - \alpha_3}a \eta_i^\bot \sim \frac{\alpha_3}{M_{\rm Pl}^2}a\eta_i^\bot\, .
\end{equation}
The spin 0 constrained variables $\delta N$ and $\d_i N_i$ are more complicated to extract, because they mix with both $\pi$ and $\d_i\eta_i$. The full calculation is straightforward but tedious, and not particularly enlightening. However the general results can be understood quite easily. The two scalar constraint equations $\delta S/\delta N$ and $(\delta S /\delta N_i)_\parallel $ are of the form
\begin{equation}
\begin{split}
H M_{\rm Pl}^2 (H\delta N + \d_i N_i) =
	& \ HM_{\rm Pl}^2 \left(H^2 \epsilon \pi + H \epsilon \dot \pi \right)\\
	&+ \alpha \left(\d_i \dot \eta_i + H \d_i\eta_i + H^3 \epsilon \pi\right)\\
	&+ \beta_2 (\d_i\eta_i + H\dot \pi + H^2 \epsilon \pi) \, ,
\end{split}
\end{equation}
where every term in the sums on both sides of the equation should be understood with order 1 coefficients. The solution is thus of the form (to linear order in fields, and for $\d_t\lesssim H$)
\begin{equation}
\{H\delta N, \d_i N_i\} = 
	H^2 \O(\epsilon\pi) + \O \left(\frac{\alpha}{M_{\rm Pl}^2}\d_i\eta_i \right)
	+ \O \left(\frac{\beta_2}{HM_{\rm Pl}^2}\d_i\eta_i \right) 
	+ H\O \left(\frac{\beta_2}{HM_{\rm Pl}^2}\dot\pi \right) \, ,
\end{equation}
which agrees with the result of \cite{Cheung:2007sv} when $\alpha,\beta_2 = 0$. Plugging this back into the action gives 
\begin{equation*}
\begin{split}
\mathcal L \sim & \ 
	c \left[ (\d_\mu\pi)^2 + \left\{\epsilon + (\beta_2^c)^2\frac{\alpha^2}{M_{\rm Pl}^4}\right\} H^2\pi^2\right]
	+ \alpha \left[(\d_\mu\eta_i)^2 - 2H^2 \eta_i^2 + \left\{\frac{\alpha \d^2 }{M_{\rm Pl}^2}+ (\beta_2^c)^2\epsilon H^2 \right\}\eta_i^2 \right] \\
	& + \beta_2 \left[\dot \pi \d\eta + \pi \frac{1}{M_{\rm Pl}^2}\left\{(c+ H\beta_2)(\frac{\alpha}{\beta}\d + 1)\right\}\eta\right]
\end{split}
\end{equation*}
where the terms in curly brackets are corrections due to mixing. Requiring these corrections to be small at energies of order $H$ leads to the constraints \eqref{DL2}.

\section{Dispersion Relations for General Mixings }
\label{app:Sta}

In this Appendix
we illustrate dispersion relations of scalar modes
in several inflationary models with mixing interactions.
As a toy model,
let us consider the following quadratic action
of two scalar fields:
\begin{align}
S=\int \frac{d\omega d^3{\bf k}}{(2\pi)^4}\left[\frac{K_\pi(\omega,k)}{2}\pi_{\omega,{\bf k}}\pi_{-\omega,{\bf -k}}
+\frac{K_\sigma(\omega,k)}2\sigma_{\omega,{\bf k}}\sigma_{-\omega,{\bf -k}}
+\beta(\omega,k)\pi_{\omega,{\bf k}}\sigma_{-\omega,{\bf -k}}
\right]\,,
\end{align}
where $K_\pi$ and $K_\sigma$ are kinetic operators
for $\pi$ and $\sigma$, respectively,
and $\beta$ is their mixing.
$\omega$ and $\bf k$
are the temporal frequency and spatial momentum,
respectively.
We here neglected time-dependence of kinetic operators and mixing interactions.
Such a simplification can generically be justified as long as
we consider modes inside the horizon.
The on-shell condition of this model can then be stated as
\begin{equation}
\text{det}\left( \begin{array}{cc}
K_\pi(\omega,k) &\beta^*(\omega,k) \\
\beta(\omega,k) &K_\sigma(\omega,k)  \end{array} \right)=0
\quad
\leftrightarrow
\quad
K_\pi K_\sigma -|\beta|^2=0
\,.
\end{equation}
In the following
let us assume that $\pi$ is massless,
and $\pi$ and $\sigma$ have the same sound speed,
\begin{align}
K_\pi=\omega^2-c_s^2k^2\,,
\quad
K_\sigma=\omega^2-c_s^2k^2-m_\sigma^2\,,
\end{align}
and illustrate dispersion relations for several types of mixing interactions.
Note that the qualitative features below
do not change even when $\pi$ and $\sigma$
have different sound speeds.

\begin{enumerate}
\item Multifield inflation type.

Let us first consider the multifield inflation type mixing.
In multifield inflation~\cite{Senatore:2010wk}
the $\sigma$ field enjoys the shift symmetry, $\sigma\to\sigma+\rm{constant}$,
so that it is massless, $m_\sigma=0$,
and the mixing interactions are generically of the form,
$\dot{\pi}\dot{\sigma}$.
Since the corresponding $\beta$ is quadratic in $\omega$ and $k$,
the dispersion relation is always linear
and the mixing interaction just modifies the propagation speeds.
For example,
when $\beta=\beta_{\rm multi}\,\omega^2$ with a real constant $\beta_{\rm multi}$,
the on-shell condition is given by
\begin{align}
\left(1-\beta_{\rm multi}^2\right)\omega^4
-2c_s^2k^2\omega^2
+c_s^4k^4=0
\quad
\leftrightarrow
\quad
\omega^2=\frac{c_s^2}{1\pm \beta_{\rm multi}}k^2\,.
\end{align}

\item Quasi-single field inflation type.

We next consider the quasi-single field inflation type model.
In quasi-single field inflation~\cite{Chen:2009zp}
there is no shift symmetry of $\sigma$,
so that it is massive, $m_\sigma\neq0$,
and there exists mixing interaction of the form $\dot{\pi}\sigma$.
For the mixing, $\beta=i\beta_{\rm quasi}\,\omega$,
with a real constant $\beta_{\rm quasi}$,
the on-shell condition can be stated as
\begin{align}
\nonumber
&\omega^4-\left(\beta_{\rm quasi}^2+m_\sigma^2+2c_s^2k^2\right)\omega^2
+c_s^2k^2(m_\sigma^2+c_s^2k^2)=0
\\[2mm]
\label{dispersion_quasi1}
\leftrightarrow\quad&
\omega^2=\frac{\beta_{\rm quasi}^2+m_\sigma^2+2c_s^2k^2\pm\sqrt{\left(\beta_{\rm quasi}^2+m_\sigma^2\right)^2+4\beta_{\rm quasi}^2c_s^2k^2}}{2}\,,
\end{align}
which contains one gapless mode and one gapped mode.
In particular,
the gapless mode has a quadratic dispersion
and the gapped mode is quite heavy
for the wavenumbers such that  $\beta_{\rm quasi}\gg c_s k\gg m_\sigma$~\cite{Assassi:2013gxa}:
\begin{align}
\label{dispersion_quasi2}
\omega^2\simeq
\frac{c_s^4k^4}{\beta_{\rm quasi}^2}\,,
\,\,
\beta_{\rm quasi}^2+2c_s^2k^2\,.
\end{align}

\item Mixing with boost breaking Goldstone boson.

Finally,
let us discuss the  case where the Goldstone boson of boost breaking are present in the theory.
Now
the massive scalar~$\sigma$
is identified with the longitudinal mode
of the boost Nambu-Goldstone mode $\eta_i$.
Because of the spin index of $\eta_i$,
this model can accommodate
mixings of the form
$\eta_i\partial_i\pi$ and $\eta_i\partial_i\dot{\pi}$,
which are linear in the spatial derivative,
without spoiling the shift symmetry of $\pi$.

Let us first consider the effect of the first type mixing, $\eta_i\partial_i\pi$.
Since the corresponding $\beta$ is given by
$\beta=\beta_{\rm boost1}k$ with a real constant $\beta_{\rm boost1}$,
the on-shell condition is
\begin{align}
\nonumber
&\omega^4-\left(m_\sigma^2+2c_s^2k^2\right)\omega^2+k^2\left(c_s^2m_\sigma^2+c_s^4k^2-\beta_{\rm boost1}^2\right)=0
\\[2mm]
\leftrightarrow\quad
&\omega^2=\frac{m_\sigma^2+2c_s^2k^2\pm\sqrt{m_\sigma^4+4\beta_{\rm boost1}^2k^2}}{2}
\,.
\end{align}
An important point is that
there appears an exponentially growing mode,
i.e., the mode with an imaginary $\omega$,
when
$k^2\left(\beta_{\rm boost1}^2-c_s^2m_\sigma^2-c_s^4k^2\right)>0$.
Therefore, the strong mixing regime can generically be unstable.
Note that, however,
the on-shell condition for the long mode, $k=0$,
is given by $\omega^2=0,m_\sigma^2$,
so that it does not experience an exponential growth.
Essentially because of that,
the stability condition at the superhorizon limit $k/a=0$
does not give any constraint on the size of the first type of mixing interaction,
as we discussed in Sec.~\ref{subsec:spin0}.

We next consider the second type of mixing interaction, $\eta_i\partial_i\dot{\pi}$.
The corresponding $\beta$
is given by $\beta=i\beta_{\rm boost2}\,\omega k$
with a real constant $\beta_{\rm boost2}$.
The on-shell condition can then be obtained
by the replacement $\beta_{\rm quasi}\to\beta_{\rm boost2}\,k$
in Eqs.~\eqref{dispersion_quasi1} and~\eqref{dispersion_quasi2}.
In particular,
in the strong mixing regime, and for wavenumbers such that $\beta_{\rm boost2}\,k\gg c_s k\gg m_\sigma$, we have
\begin{align}
\omega^2\simeq
\frac{c_s^4}{\beta_{\rm boost2}^2}k^2\,,
\,\,
\beta_{\rm boost2}^2k^2\,.
\end{align}
In contrast to the quasi-single field inflation case,
we have two modes with a linear dispersion.
As we explained in Sec.~\ref{subsec:spin0},
one mode propagates with a very small sound speed,
where as the other is quite superluminal, and we do not explore this exotic possibility. 

\end{enumerate}

\section{Concrete Form of Shape Functions}
\label{App:shape}

In this Appendix
we summarize details of the bispectrum.
Performing the integrals in~\eqref{shape_1},
we obtain the shape function
of the form,
\begin{align}
S(k_1,k_2,k_3)&=
\frac{M^2}{\bar{M}H}(\beta_2^c)^3
\Big[
I(k_1,k_2,k_3)
+I(k_2,k_3,k_1)
+I(k_3,k_1,k_2)
\Big]\,,
\end{align}
where $I(k_1,k_2,k_3)$ is given by
\begin{align}
\nonumber
I(k_1,k_2,k_3)
&=\frac{1}{16c_\pi^2r^2(1-r)^2(1+r)^2}\frac{\kappa_1\left(-\kappa_1^2+\kappa_2^2+\kappa_3^2\right)}{\kappa_2\kappa_3}
\\
\nonumber
&\quad\times
\Bigg[(2-\kappa_1)r^2+\frac{2}{\kappa_1+r\kappa_2+r\kappa_3}
-\frac{2r\left(1+\kappa_1+r\kappa_2+r\kappa_3\right)}{(\kappa_1+r\kappa_2+\kappa_3)(\kappa_1+\kappa_2+r\kappa_3)}
\\
&\qquad\quad
+\frac{r\kappa_1}{(\kappa_1+r\kappa_2+\kappa_3)^2}
+\frac{r\kappa_1}{(\kappa_1+\kappa_2+r\kappa_3)^2}
-\frac{\kappa_1}{(\kappa_1+r\kappa_2+r\kappa_3)^2}
\Bigg]\,.
\end{align}
Here $\kappa_i=k_i/(k_1+k_2+k_3)$ and $r=c_\sigma/c_\pi$.
It is useful to introduce
concrete expressions of the shape function itself for typical values of $r$.
First,
when $\pi$ and $\sigma$ have the same sound speed ($r=1$),
\begin{align}
\nonumber
S(k_1,k_2,k_3)&=
\frac{M^2}{\bar{M}H}
\frac{(\beta_2^c)^3}{64c_\pi^2}
\Bigg(
4-4\sum_i\kappa_i^2
-4\sum_{i\neq j}\kappa_i\kappa_j^2
\\
&\qquad\qquad\qquad\quad
+\frac{
-\sum_i\kappa_i^2
+\sum_{i\neq j}\kappa_i\kappa_j^2
+4\sum_{i<j}\kappa_i^2\kappa_j^2
}{\kappa_1\kappa_2\kappa_3}
\Bigg)\,.
\end{align}
Next, when $c_\sigma\ll c_\pi$ ($r\ll1$),
we have
\begin{align}
S(k_1,k_2,k_3)&=
\frac{M^2}{\bar{M}H}
\frac{(\beta_2^c)^3}{16c_\sigma^2}
\frac{\sum_{i\neq j}\kappa_i\kappa_j^2-\sum_{i}\kappa_i^3}{\kappa_1\kappa_2\kappa_3}\,.
\end{align}

\bibliographystyle{h-physrev}
\bibliography{BoostEFToI_JournalSub}{}

\end{document}